\documentclass[11pt]{article} 
\usepackage[utf8]{inputenc}
\usepackage{multirow}
\usepackage{amsfonts}
\usepackage{amsmath}
\usepackage{cancel}
\usepackage{amssymb}
\usepackage{amsthm}
\usepackage{booktabs, array}
\newcolumntype{M}{>{$}c<{$}}  
\usepackage[export]{adjustbox}
\usepackage[english]{babel}
\usepackage{tikz}
\usetikzlibrary{shapes.misc,decorations.markings,arrows,decorations.pathmorphing,backgrounds,matrix,patterns,positioning,decorations.pathreplacing,calc}
\usetikzlibrary{arrows.meta}
\tikzset{cross/.style={cross out, draw=blue, minimum size=2*(#1-\pgflinewidth), inner sep=0pt, outer sep=0pt}, cross/.default={1pt}}
\tikzset{
  tightzigzag/.style={decorate,
    decoration={
      zigzag,  segment length=6pt,  %
      amplitude=1.2pt,     %
      pre=curveto, post=curveto}}}

\makeatletter
\usepackage{comment}
\usepackage{float}
\usepackage{caption}
\usepackage{xcolor}
\definecolor{dgreen}{rgb}{0,0.5,0}
\definecolor{darkblue}{rgb}{0,0,0.6}
\definecolor{lblue}{rgb}{0.5,0.5,1.0}
\definecolor{purple}{rgb}{0.4,.2,0.7}
\definecolor{ggray}{rgb}{0.85,.85,0.85}
\definecolor{orange}{rgb}{1,.5,0}
\def\red{\color{red}}

\usepackage[margin = 2.5cm]{geometry}
\usepackage{graphicx}
\usepackage[hyperfootnotes = false, colorlinks = true, linkcolor = blue, citecolor = purple]{hyperref}
\usepackage{subcaption}
    \usepackage{blkarray}

\def\ym{\mathrm{YM}}
\def\la{\label}
\def\tr{\mathrm{tr}}

\usepackage[most]{tcolorbox}
\newtcolorbox[auto counter]{exercise}[2][]{
  enhanced,
  breakable,
  colback=gray!10,           %
  colframe=gray!70!black,    %
  fonttitle=\bfseries,
  title=Exercise~\thetcbcounter: #2,  %
  coltitle=black,
  attach boxed title to top left={
    yshift=-2pt, xshift=4pt
  },
  boxed title style={
    colback=gray!30,
  },
  #1                        %
}

\newtcolorbox[auto counter]{boxcomment}[2][]{
  enhanced,
  breakable,
  colback=pink!30,           %
  colframe=gray!70!black,    %
  fonttitle=\bfseries,
  title=Further reading~\thetcbcounter: #2,  %
  coltitle=black,
  attach boxed title to top left={
    yshift=-2pt, xshift=4pt
  },
  boxed title style={
    colback=pink!30,
  },
  #1                        %
}

\newtcolorbox[auto counter]{boxc}[2][]{
  enhanced,
  breakable,
  colback=gray!10,           %
  colframe=gray!70!black,    %
  fonttitle=\bfseries,
  title= #2,  %
  coltitle=black,
  attach boxed title to top left={
    yshift=-2pt, xshift=4pt
  },
  boxed title style={
    colback=gray!30,
  },
  #1                        %
}

\newcommand{\bes}{\begin{equation} \begin{split} }	
	\newcommand{\ees}{\end{split} \end{equation} }
	
	\renewcommand{\i}{\mathrm{i}}

	\newcommand{\RA}{\Rightarrow}

	\newcommand{\pd}{\partial}

	\newcommand{\inv}{^{-1}}
	
	\newcommand{\hf}{\tfrac{1}{2}}
	\newcommand{\qrt}{\frac{1}{4}}
	\usepackage{physics}
	\def\tfd{\mathrm{TFD}}

\usepackage{amsfonts}
\usepackage{amsmath}
\usepackage{amssymb}
\usepackage{braket}
\usepackage{cite}
\usepackage{enumitem}
\usepackage{float}
\usepackage{graphicx}
\usepackage{mathrsfs}
\usepackage{setspace}
\usepackage{physics}
\usepackage{subcaption}
\usepackage{titlesec}
\usepackage{comment}

\def\hf{\tfrac{1}{2}}
\def\la{\label}
\def\nref#1{(\ref{#1})}

\usepackage[margin = 2.5cm]{geometry}
\usepackage{color}
\definecolor{darkblue}{rgb}{0,0,0.6}
\definecolor{purple}{rgb}{0.4,.2,0.7}
\definecolor{darkgreen}{rgb}{0,0.5,0}

\renewcommand{\i}{\mathrm{i}}
\renewcommand{\d}{\mathrm{d}}
\newcommand{\ads}{\mathrm{AdS}}

\begin{document}

\thispagestyle{empty}
\begin{center}
    ~\vspace{5mm}

     {\LARGE \bf TASI lectures on Matrix Theory from a modern viewpoint}

   \vspace{0.5in}
     
   {Henry W. Lin$^{a,b}$ }

    \vspace{0.5in}

   ~
   \\
   {$^a$ Leinweber Institute for Theoretical Physics, Stanford University, Stanford, CA 94305, USA}  \\
   {$^b$ Jadwin Hall, Princeton University, Princeton, NJ 08540, USA}
                
    \vspace{0.5in}

    \vspace{0.5in}
    
\end{center}

\vspace{0.5in}

\begin{abstract}
These notes review the D0-brane or Banks-Fischler-Shenker-Susskind (BFSS) matrix quantum mechanics from a post-AdS/CFT perspective. We start from the decoupling argument for D0-branes and discuss the gravity dual in the 't Hooft regime, before extrapolating to strong coupling. In the second part of these notes, we review the matrix bootstrap method and its application to the D0-brane quantum mechanics.

\end{abstract}

\vspace{1in}

\pagebreak

\setcounter{tocdepth}{3}

\tableofcontents

\section{Introduction}

The Banks-Fischler-Shenker-Susskind (BFSS) conjecture \cite{Banks:1996vh} was proposed in 1996. It is probably older than the median TASI audience member. Why should we revisit it now? On the one hand, the motivation for studying the model is as strong as ever:
\begin{enumerate}
    \item It provides a non-perturbative definition of the M-theory scattering matrix. When BFSS made their conjecture, there was no other candidate for the non-perturbative S-matrix. However, post-AdS/CFT we may also define the M-theory S-matrix by taking the flat space limit; see \ref{mtheory_approach} for more on this. So a sharper version of the conjecture is that all of these different definitions in fact compute the same S-matrix.
    \item It contains black holes, including ones that are well described by Einstein gravity.
    \item It is ``just'' a quantum mechanical model (as opposed to a field theory), and therefore should be somewhat easier to simulate via classical or quantum \cite{Maldacena:2023acv} algorithms. Conceptually, it is a distillation\footnote{Perhaps the IKKT model \cite{Imamura:1997ss} is an even more potent distillation, where all of spacetime emerges from just an integral.} of the holographic Mystery, since all of space emerges from just several matrices sitting in a ``quantum dot.'' Furthermore, in the scattering setup relevant for the BFSS conjecture, spacetime emerges in a way that seems qualitatively different than, say, in the flat space limit of AdS/CFT. 
\end{enumerate}
We have two main advantages over physicists in the 90s: 
\begin{enumerate}
    \item[{\color{blue} (1)}] We understand other examples of gauge/gravity duality in much more detail; presumably this collective experience can inspire progress\footnote{Besides the tremendous amount of progress in AdS/CFT, I would like to highlight some recent progress \cite{Hartnoll:2024csr, Komatsu:2024bop, Komatsu:2024ydh} in a lower dimensional cousin of BFSS, the IKKT model. Recently a massive deformation of this model has been studied that is a bit analogous to the massive deformation of BFSS, the so-called pp wave or Berenstein-Maldacena-Nastase (BMN) model \cite{Berenstein:2002jq}.}.
    \item[{\color{blue} (2)}] We have better numerics, both algorithms and machines. We are perhaps a little less intimidated by strong coupling and large $N$. 
\end{enumerate}
In part one of these lectures, I will review our understanding from the gauge/gravity point of view, taking advantage of {\color{blue} (1)}. I will not review the original arguments for BFSS which involve thinking about the infinite momentum frame, see \cite{bigatti1997reviewmatrixtheory, Polchinski:1999br, Taylor:2001vb, MaldacenaStrings} for reviews which cover this. Instead I will try to spell out the logic from the point of view of someone who pretty much takes AdS/CFT for granted. %

In part two, we will review some recent numerical approaches to matrix theories {\color{blue} (2)}. This will be {\it hugely biased} towards the bootstrap approach, although lattice Monte Carlo is a very important method \cite{Anagnostopoulos:2007fw, Hanada:2008ez, Catterall:2008yz, Catterall:2009xn, Filev:2015hia, Rinaldi:2021jbg, Pateloudis:2022ijr, Bodendorfer:2024egw}, see also \cite{Balthazar:2016utu, Han:2019wue, Koch:2021yeb,  Mathaba:2023non, Bodendorfer:2024egw} for other recent numerical approaches to related models. 
Some simple exercises are also included; the reader is encouraged to do them!

\section{Review of D0-brane holography \label{sec:review}}

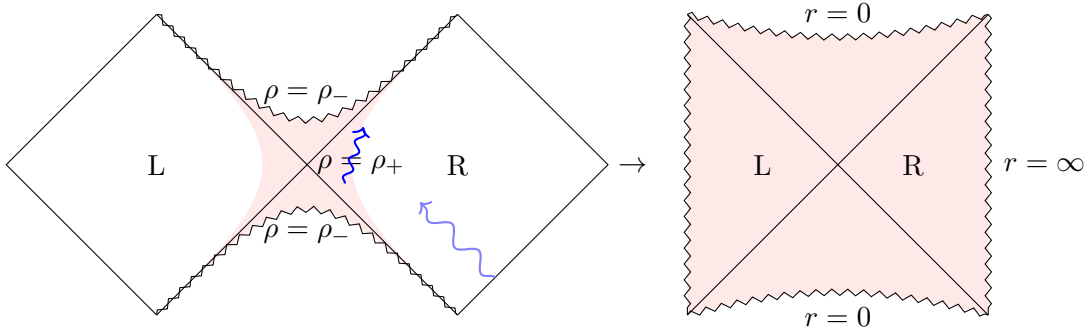
\begin{figure}
\centering
\begin{tikzpicture}[scale=0.5,baseline={([yshift=-.55ex]current bounding box.center)}]
\node (I)    at ( 4,0) {R};  %
\node[right] at (0,0) {$\rho = \rho_+$};
\node (II)   at (-4,0) {L}; %
\path  
  (II) +(90:4)  coordinate  (IItop)
       +(-90:4) coordinate (IIbot)
       +(0:4)   coordinate                  (IIright);
\draw     (IItop) -- (IIright) -- (IIbot) ;
\path 
   (I) +(90:4)  coordinate (Itop)
       +(-90:4) coordinate (Ibot)
       +(180:4) coordinate (Ileft);
\draw  (Itop) -- (Ileft) -- (Ibot); 
\draw[tightzigzag] (IItop) to[bend right=45, looseness=1.7] node[midway, above, inner sep=2mm] {$\rho = \rho_- $} (Itop);
\draw (IItop) --(-8,0) -- (IIbot);
\draw[decorate,decoration] (Itop)  --(8,0) --  (Ibot);
\draw[tightzigzag] (IIbot) to[bend left=45, looseness=1.7] node[midway, below, inner sep=2mm] {$\rho = \rho_- $} (Ibot)  ;
\draw[thick, lblue, decorate, decoration={snake, amplitude=1mm, segment length=5mm}, ->]    (5,-3) -- (3,-1);
\begin{pgfonlayer}{background}
  \path[fill=red!25,opacity=.35]
        (IItop)
        to[bend right=45,looseness=1.7] (Itop)
        to[bend right=45,looseness=1.7] (Ibot)
        to[bend right=45,looseness=1.7] (IIbot)
        to[bend right=45,looseness=1.7] (IItop)
        -- cycle;
\end{pgfonlayer}
\draw[thick, blue, decorate, decoration={snake, amplitude=0.5mm, segment length=2.5mm}, ->]    (1,-0.5) -- (1.5,1);
\end{tikzpicture} $\rightarrow$
\begin{tikzpicture}[scale=0.5,baseline={([yshift=-.55ex]current bounding box.center)}]
\node (I)    at ( 4,0) {};  %
\node at (2,0) {R};
\node at (-2,0) {L};
\node (II)   at (-4,0) {}; %
\node (III)  at (0, 2.5) {};
\node (IV)   at (0,-2.5) {};
\path  
  (II) +(90:4)  coordinate  (IItop)
       +(-90:4) coordinate (IIbot)
       +(0:4)   coordinate                  (IIright);
\draw     (IItop) -- (IIright) -- (IIbot) ;
\path 
   (I) +(90:4)  coordinate (Itop)
       +(-90:4) coordinate (Ibot)
       +(180:4) coordinate (Ileft);
\draw  (Itop) -- (Ileft) -- (Ibot); 
\draw[tightzigzag] (IItop) to[bend right=15] node[midway, above, inner sep=2mm] {$r=0$} (Itop);
\draw[tightzigzag] (IItop) -- (IIbot)
    node[midway, left, inner sep=2mm] {};
\draw[tightzigzag] (Itop) -- (Ibot)
        node[midway, right, inner sep=2mm] {$r=\infty$};   
\draw[tightzigzag] (IIbot) to[bend left=15] node[midway, below, inner sep=2mm] {$r=0$} (Ibot)  ;
\begin{pgfonlayer}{background}
  \path[fill=red!25,opacity=.35]
        (IItop) to[bend right=15] (Itop) to (Ibot) to[bend right=15] (IIbot) to (IItop)
        -- cycle;
\end{pgfonlayer}
\end{tikzpicture}
\caption{\la{penrose_decoupling} {\it Left}: the Penrose diagram of the near-extremal D0-brane black hole. The near-horizon region is shaded in {\color{pink} pink}. We also show two excitations that survive the decoupling limit: one in the near-horizon region, and one in the flat space region. {\it Right}: after taking the decoupling limit, we are left with a geometry whose Penrose diagram resembles that of an AdS black brane. In addition to the black hole singularity, there is a boundary singularity near $r=\infty$.}
\end{figure}

Let us review the gravity dual of SYM in $0+1$ spacetime dimensions, following the logic of \cite{Maldacena:1997re, Itzhaki:1998dd}, (see also \cite{Polchinski:1999br}). 

\subsection{D0 black hole}
We start by recalling the D0 black hole solution. Since D0 branes carry Ramond-Ramond charge, this is a charged black hole.
We use conventions where the relevant part of the Type II supergravity action is
\begin{align}\label{actionII}
   I =  \frac{1}{(2\pi)^7 \ell_s^8} \int \d^{10} x \sqrt{g} \left[e^{-2 \phi} (R+4 (\nabla \phi)^2) -\frac{1}{4} F_{\mu \nu }^2\right]
\end{align}
From the spacetime point of view, the D0 branes source the (string-frame) metric, dilaton, and Ramond-Ramond fields \cite{Horowitz:1991cd, Duff:1993ye, Klebanov:1996un, Duff:1996hp}:
\begin{align} \label{string_frame}
    \d s^2 &= - H^{-1/2}(r) f(r)  \d t^2 + H^{1/2}  (f^{-1}(r) \d r^2 + r^2 \d \Omega_8^2),\\
     e^{-2 \phi} &= g_\mathrm{s}^{-2} H^{-3/2},    \quad A_0  = \frac{1}{g_s}(H^{-1}-1) \coth \alpha ,\\
     H &= 1 + \frac{r_0^7 \sinh^2 \alpha }{r^7}, \quad  %
    f = 1 - \frac{r_0^7}{r^7} \label{h_eq}
\end{align}
We can exchange the parameters $r_0, \alpha$ for the mass and charge $(M,Q)$ of the black hole:
\begin{align} \label{massAndCharge}
    M &= \frac{ (2\pi)^2}{  7 d_0 } \frac{ r_0^7}{   g_s^2 \ell_s^8 } (7 \cosh^2 \alpha +1), \quad  N = \frac{ (2\pi)^2}{  d_0  } \frac{r_0^7}{g_s\ell_s^7 } (\cosh \alpha \sinh \alpha),\\
    \frac{M}{N} &\ge \frac{1}{g_s \ell_s} , \quad d_0  = 60 \pi^3 (2\pi)^2 = 240 \pi^5\label{BPSbound}.%
\end{align}
The formula \eqref{BPSbound} is the gravitational manifestation of the BPS bound.
The temperature of the black hole is:
\begin{align}
    T = \frac{7}{4\pi r_0 \cosh \alpha } %
    \label{temp}
\end{align}
We see that in the extremal limit $T \to 0$ at fixed $N$, we recover the familiar fact that a D0 brane has mass $1/g_s $ in string units.

How do we get this solution? We will take a somewhat scenic route. Recall that Type IIA can be viewed as M-theory in 11d compactified on a small circle of radius $R =  g_s \ell_s$. The dilaton is interpreted as the size of the circle, and the RR 1-form comes from the off-diagonal components of the metric. This immediately suggests that we should search for an 11d solution which is pure metric. The most naive guess is simply the Schwarzschild black hole with $d=11$:
\begin{align} 
    \left( \frac{\d s^2}{\ell_p^2} \right)_\text{black hole} &=- f(r) \d t^2+f(r)\inv {\d r^2}+r^2 \d \Omega_{d-2}^2, \quad f=1-\frac{r_0^{d-3}}{r^{d-3}}.
\end{align}
This is not a bad guess, but there are two problems. First, we want a solution that is uniform in the 11th dimension, so that we can do dimensional reduction. Second, we want the solution to carry a large amount of RR charge. There is a simple fix to both issues. First, we consider the black string in 11d. This means we take the Schwarzschild solution in $d=10$ and add a dimension: 
 \begin{align} \label{blackString}
    \left( \frac{\d s^2}{\ell_p^2} \right)_\text{black string} &=-f \d t^2+ f(r)\inv {\d r^2}+r^2 \d \Omega_8^2+\d x_{11}^2, \quad f=1-\frac{r_0^7}{r^7} .
\end{align}
This fixes objection (1). To address objection (2), let's perform a Lorentz boost:
\begin{align} \label{lorentzBoost}
 \binom{t}{x_{11} } \to \binom{t' }{x_{11}'}=\left(\begin{array}{cc}
\cosh \alpha & \sinh \alpha \\
\sinh \alpha & \cosh \alpha
\end{array}\right)\binom{t}{ x_{11} } .
\end{align}
For large boost parameter, we now have a candidate solution which is uniform in the 11th dimension, and contains lots of Kaluza--Klein momentum:
\begin{align} \label{boostedString}
\d s_{11}^2 &= H^{-1}\left(-f \,  \d t^2\right)+H\left[\d z+\left(H^{-1}-1\right) \operatorname{coth} \alpha \, \d t\right]^2+f^{-1} \d r^2+r^2 \d \Omega_8^2,%
\end{align}
where $H(r)$ is defined as before \eqref{h_eq}.
\begin{exercise}{Reduction of 11d black string \label{boostedStringReduce}}
\begin{enumerate}
    \item Perform the change of coordinates \eqref{lorentzBoost} and check that we arrive at \eqref{boostedString}. 
    \item Check that the solution \eqref{boostedString} agrees with the type IIA reduction \eqref{string_frame}, via $\d s_{11}^2=e^{4 \phi / 3}\left(\d z+A_\mu \d x^\mu\right)^2+e^{-2 \phi / 3} \d s_{10}^2 .$
\end{enumerate}
\end{exercise}

\begin{exercise}{Gregory Laflamme Instability \label{GLexercise}}
Give a poor man's argument for the dynamical instability \cite{Gregory:1993vy} of the black string: in $d+1$ spacetime dimensions, compute the entropy of a black string of length $L$ and compare it to the entropy of a black hole with the same mass $M$. At what scale do the entropies become comparable?
\end{exercise}

\subsection{Decoupling limit}
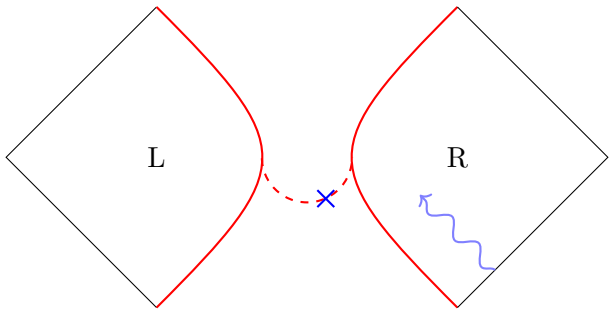
\begin{figure}
\centering
\begin{tikzpicture}[scale=0.5,baseline={([yshift=-.55ex]current bounding box.center)}]
\node (I)    at ( 4,0) {R};  %
\node (II)   at (-4,0) {L}; %
\path  
  (II) +(90:4)  coordinate  (IItop)
       +(-90:4) coordinate (IIbot)
       +(0:4)   coordinate (IIright);
\path 
   (I) +(90:4)  coordinate (Itop)
       +(-90:4) coordinate (Ibot)
       +(180:4) coordinate (Ileft);
\draw (IItop) --(-8,0) -- (IIbot);
\draw[red, thick] (IItop) to[bend left=45,looseness=1.7] (IIbot);
\draw[red, thick] (Itop) to[bend right=45,looseness=1.7] (Ibot);
\draw (Itop)  --(8,0) --  (Ibot);
\draw[thick, lblue, decorate, decoration={snake, amplitude=1mm, segment length=5mm}, ->]    (5,-3) -- (3,-1);
\begin{scope}
    \clip (-1.5,0) rectangle (1.5,-1.5);
    \draw[dashed,red,thick] (0,0) circle(1.2);
\end{scope}
    \node [cross=4pt,thick] at (0.5,-1.1) {};
\end{tikzpicture}
\caption{\label{decoupling} In the decoupling limit, we can imagine replacing the near horizon region with the super Yang Mills ``{\color{red} quantum dot}.'' In principle, we are left with two copies of the D0 brane quantum mechanics, entangled in the thermofield double state $\ket{\tfd}$. We also show a low energy, massless mode propagating in the flat space region that could be absorbed by the D0 brane. }
\end{figure}

Now, following \cite{Maldacena:1997re, Itzhaki:1998dd} we take the decoupling limit: 
\begin{align}
 g_s \ll 1 \quad \text{holding fixed:} \quad  \ell_s \Delta  E  \sim g_s^{1/3} , \quad N.
\end{align}
This means we consider type IIA string theory in asymptotically flat space with $N$ D0 branes and take the $g_s \to 0$ limit. We focus on the low-energy states in the Hilbert space with energy above extremality\footnote{The BPS bound ensures there are no states below extremality.} $\Delta E \sim g_s^{1/3} /\ell_s$.

The obvious states are massless modes (e.g. supergravitons) with very low energy. But there are also excitations of the D0 branes. The D0 brane theory is described by super Yang Mills with SU($N$) gauge group $+$ higher derivative interactions:
\begin{align}
    S \sim \frac{1}{g^2_\ym} \int  \d t  \, \left[ \frac12  \Tr \dot{X}^2 + \frac{1}{4} \Tr [X_I ,X_J]^2 + \text{fermions}+ \text{higher derivatives} \right] 
\end{align}
Since we are only considering very low energy states (much lower than the string scale), we can ignore the higher derivative interactions. Then the only dimensionful scale in the SYM theory is given by the coupling\footnote{The SYM theory is conventionally defined so that the Hamiltonian measures the energy above extremality. E.g., we subtract off the rest mass of the D0 branes $N/(g_s \ell_s)$.}. For SYM theory in $p+1$ dimensions, the Yang-Mills coupling has units of
\begin{align}
    [g^2_\ym ] = \text{energy}^{3-p} , \quad g^2_\ym =  \frac{g_s}{(2\pi)^{2-p} \ell_s^{3-p}}.
\end{align}
So from the SYM description, the natural energy scale is $\Delta E \sim (g^2_\ym)^{1/3} \sim g_s^{1/3} / \ell_s $ which is being held fixed in this limit.

Now let us analyze the same set of states from the bulk spacetime perspective. We are instructed to consider a black hole with fixed charge $N$. Using \eqref{massAndCharge}, the energy above extremality in the decoupling limit is
\begin{align}
    \Delta E \equiv M-\frac{N}{g_s\ell_s}
    =\frac{9}{14}\frac{(2\pi)^2}{d_0}
    \frac{r_0^7}{g_s^2\ell_s^8} .
    \label{energyAboveExtremality}
\end{align}
This should scale as $\Delta E\sim g_s^{1/3}/\ell_s$. It follows that
\begin{align}
r_0/\ell_s \sim g_s^{1/3}.
\end{align}
Equivalently, in terms of the black-hole parameters, we take $r_0/\ell_s\sim g_s^{1/3}$ and $\alpha\to\infty$, holding $r_0/(\ell_s g_s^{1/3})$ and $g_s^{2/3}e^\alpha$ fixed.
So the parameter $r_0$ is small in string units, which tells us that the black hole is near extremality from the flat space point of view.
Alternatively, we can demand that $\beta /\ell_s \sim  g_s^{-1/3}$. Then the scaling of $r_0$ follows from \eqref{temp}. We sketch the Penrose diagram in the near-extremal limit in Figure \ref{penrose_decoupling}. The black hole singularity formally becomes null in the extremal limit\footnote{In the decoupled geometry, it is a causal feature of the Penrose diagram that the singularity bends down. But the precise shape of the singularity depends on the conformal transformation we use to draw the diagram. In the decoupled geometry (with flat space asymptotics) the shape of the singularity is arbitrary but in the strict extremal limit the singularity becomes null. I thank Douglas Stanford for some discussion about this point, see also \cite{Fidkowski_2004}.}.

So far we have argued that the black hole geometry (near extremality) survives the decoupling limit. But we would like to know what kinds of perturbations to this background also survive. Consider a particle that is located in the near horizon region, e.g., at $r/\ell_s \sim g_s^{1/3}$. More precisely, let us introduce a new radial coordinate:
\begin{align} \label{rhocoord}
\rho = \frac{r}{\alpha'} \frac{1}{(d_0 g_\mathrm{YM}^2 N)^{1/3}}.
\end{align}
We consider a particle at a fixed $\rho$ in the decoupling limit. Now let us calculate the energy of the particle relative to the flat space region. We must account for the redshift factor $\sqrt{g_{tt}}  = H^{-1/4}$. This means that the energy as measured by the boundary system (or as seen from asymptotically flat space) is quite a bit lower:
\begin{align}
    E =  H^{-1/4} E_p \sim (g_s N)^{1/3} \rho^{7/4} E_p 
\end{align}
We see that even if the particle has energy $E_p \sim 1/\ell_s $ (e.g. an excited string state), it will survive the decoupling limit. This means that a generic object in string theory will survive the decoupling limit if it resides in the near horizon region\footnote{One could also ask about D-branes which have parametrically larger masses than $1/\ell_s$. But note that the effective string coupling in the throat is finite, so these objects are not parametrically heavier in terms of $1/g_s$ where $g_s$ is the string coupling in the flat space region. Hence they too survive decoupling.}.

Based on these considerations, it is convenient to define a new dimensionless time coordinate $\tau = \left( d_0 g^2_{\mathrm{YM}}  N \right) ^{1/3} t $. The energy $\mathcal{E}$ conjugate to $\tau$ is the dimensionless energy measured in units of the 't Hooft coupling (which survives the decoupling limit), e.g., $\mathcal{E} = H/(d_0 g^2_\ym N)^{1/3}$. In these coordinates, the near horizon region is
\begin{align}
    \frac{\d s^2}{\alpha'} &=-\rho^{7/2} f(\rho)\, \d \tau^2+\frac{\d \rho^2}{\rho^{7/2} f(\rho)}+ \rho^{-3/2}\d\Omega_8^2 ,
     \label{eq:metric}\\
e^{-\phi} &= \frac{d_0 N  }{(2\pi)^2}\rho^{21/4},%
   \quad A_0 = \ell_s \frac{d_0 N}{(2\pi)^2} \rho^7 %
 , \quad d_0 = 240 \pi^5 %
\end{align}
Note that even though we sent $g_s\to 0$ the effective string coupling is finite in the decoupling limit (e.g., independent of the flat space $g_s$). This has two implications. It means that excitations that are heavier than massive strings, e.g., D-branes, also survive the decoupling limit\footnote{For some work on D-branes in the decoupled D$p$-brane backgrounds, see \cite{Das:2000ab, Batra:2025ivy}. These works discuss operators in the boundary theory that create giant gravitons, mimicking to some extent the discussion of giant graviton operators in AdS/CFT.}. Second, to make the effective string coupling $\sim 1/N$ small (so that we can trust the semi-classical IIA approximation), we must take $N \to \infty$ and work in the 't Hooft limit. Let us emphasize that whether or not the supergravity description is reliable is secondary, it is not crucial to the decoupling argument.

To summarize, according to the gravity solution, we expect that there are two kinds of excitations in the $g_s \to 0$, low energy limit. There are the low energy massless excitations of weakly coupled IIA string theory in flat space. And there are generic excitations (massless or massive excitations) of the bulk theory in the near-horizon region. So the Hilbert space approximately factorizes:
\begin{align} \label{dec1}
    \mathcal{H}_\text{IIA string theory} \approx \mathcal{H}_\text{flat space} \otimes \mathcal{H}_\text{near horizon region}
\end{align}
On the other hand, from the worldvolume (worldline in the D0 case) perspective: %
\begin{align} \label{dec2}
    \mathcal{H}_\text{IIA string theory} \approx \mathcal{H}_\text{flat space} \otimes \mathcal{H}_\text{SYM quantum mechanics}
\end{align}
We have illustrated this in Figure~\ref{decoupling}. Cancelling the Hilbert space $\mathcal{H}_\text{flat space}$ in both equations \eqref{dec1} and \eqref{dec2} leads us to identify $\mathcal{H}_\text{near horizon region} = \mathcal{H}_\text{SYM quantum mechanics}$. %

\begin{exercise}{Black hole in a box \label{blackholebox}}
    It is often said that AdS is a convenient way to put black holes in a box. Show that the D0 brane geometry is also a ``box'' in the sense that particles cannot escape to infinity. Show that the only exception to this rule is a D0 brane itself; the D0 brane experiences a gravitational force which is essentially equal and opposite of the electric force. (The BPS bound forbids an object with a greater charge to mass ratio.) This means that the black hole can only Hawking evaporate into D0 branes. Estimate the lifetime of the black hole. (For a detailed estimate, see \cite{Lin:2014wka}.)
\end{exercise}

\begin{exercise}{D0 brane thermodynamics \label{thermodynamicsEx}}
\begin{enumerate}
\item Work out the black hole thermodynamics of the above metric. Start by computing the temperature of the metric \eqref{eq:metric}. Then compute $S=\frac{A}{4G_N}$, where $A$ is the area of the horizon in Einstein frame, and express it in terms of the temperature. Finally use this to derive the energy of the black hole as a function of temperature. 
\item Consider a higher derivative curvature correction $\sim \alpha'^3 \int \sqrt{g} \, e^{-2 \phi} R^4 $. How does this affect the thermodynamics? Use this to give a parametric upper bound on the temperature for which the supergravity analysis is valid. Even at low temperatures, argue that the Type IIA solution is valid for $r/\ell_p \ll N^{1/3}$.
\item When $e^\phi \sim 1$, the 10d solution should be replaced by the 11d black string. Work out the temperature scale where this occurs. Argue that even at these low temperatures, we can still use the 10d solution as long as  $r/\ell_p \gg N^{1/7}$.
At even lower temperatures, the results of exercise \ref{GLexercise} imply that the 11d black string will transition into an 11d black hole. Use these results to give a more complete picture of the thermodynamics at very low temperatures \cite{Horowitz:1997fr, Itzhaki:1998dd}. For recent work, see \cite{Dias:2024vsc}.
\end{enumerate}
\end{exercise}

How big is the ``throat'' in the decoupling limit? At low temperatures, the proper length from the horizon to the ``boundary'' or flat space region goes like $\Delta s \sim \ell_s \tilde \beta ^{-3/10}$ where $\tilde{\beta} = \beta (d_0 g^2_\ym N)$ is the dimensionless inverse temperature in 't Hooft units.
We see that for order 1 temperatures, the entire region is string scale. This suggests that the metric is unreliable at these temperatures due to $\alpha'$ corrections. Indeed, in exercise \ref{thermodynamicsEx}, you will check that the geometry is only trustworthy at low temperatures.

\begin{boxcomment}{Other Dp brane geometries}
	Study the other Dp brane geometries for $p \le 4$ in the decoupling limit \cite{Itzhaki:1998dd}, and understand their IR behavior. \begin{enumerate}
	    \item For $p=1$, the effective string coupling grows in the radial direction. Since we have a IIB solution, we should use S-duality to find a more convenient description of the IR. Then $\phi \to -\phi$ and the coupling starts to decrease towards the IR. (In exercise \ref{matrixString} you are asked to study the holographic dual in more detail.) On the boundary, the IR behavior is ``matrix string theory'' \cite{Dijkgraaf:1997vv}, whose effective description is a free symmetric orbifold CFT perturbed by irrelevant operators. 
        There is an interesting conjecture \cite{Dijkgraaf:1997vv, Arutyunov:1997gi} that directly connects conformal perturbation theory in the leading irrelevant operator (a supersymmetric twist operator) with the genus expansion in the worldsheet approach to Type IIA scattering amplitudes. In the matrix string conjecture the string coupling is identified with $1/(g_\ym R)$, where the 1+1D SYM theory is compactified on a spatial circle of radius $R$. 
        \item For $p=2$, there is an M-theory description of the IR in terms of the M2 brane solution. The boundary description is the ABJM CFT. 
        \item $p=3$ is the familiar case of $\mathcal{N} = 4$ SYM. Note in the decoupling limit we set $g_s = $ constant. 
        \item For $p=4$, the SYM interaction is {\it irrelevant.} We view this theory as the effective field theory of M5 branes wrapped on a circle. Note that in the decoupling limit, we send $g_s \to \infty$. The UV is strongly coupled, and the bulk geometry becomes $\ads_7 \times S_4$.
        \item The $p=-1$ case is known as IKKT \cite{Ishibashi:1996xs}. A mass deformation of this model has been the subject of some interesting new developments, see \cite{Hartnoll:2024csr, Komatsu:2024bop, Komatsu:2024ydh}.
	\end{enumerate} 
\end{boxcomment}

\begin{exercise}{The holographic dual of 1+1D SYM \label{matrixString}}
In this exercise, you are asked to work out the holographic dual of maximally supersymmetric Yang--Mills theory. Start with the Type IIB extremal black D1 brane solution:
    \begin{align}
\frac{\d s^2}{\alpha'} & =\frac{U^3}{g_{\ym} \sqrt{2^6 \pi^3 N}} d x_{\|}^2+\frac{g_{Y M} \sqrt{2^6 \pi^3 N}}{U^3} d U^2+g_\ym \frac{\sqrt{2^6 \pi^3 N}}{U} d \Omega_8^2  \\
e^\phi & =\left(\frac{g_\ym^6 2^8 \pi^5 N}{U^6}\right)^{1 / 2}, \quad A_0 = -\frac{1}{2} \left(\frac{g_\ym^6 2^8 \pi^5 N}{U^6}\right)^{1 / 2}
\end{align}

In the IR, the dilaton grows and we should use S-duality. Show that this gives the extremal F1 solution:
\begin{align}
\d s^2_{F1}
&= \frac{\alpha'}{(2\pi)g_{\ym}^{2}}
\big[\,h(U)\,(-\d t^2+\d x_1^2)+\d U^2+U^2 \d\Omega_7^2\,\big], \label{F1metric}\\
B_{t1}&=h(U),\qquad
e^{\phi}=\frac{1}{(2\pi)\,g_\ym^{2}}\,
\frac{U^{3}}{\sqrt{d_1\,g_{\ym}^{2}\,N}}\,. \label{F1fields}
\end{align}
Now we compactify the SYM theory on a spatial circle $x_1 \sim x_1 + 2\pi R$. We may introduce \(\theta\)-coordinates, so that \(g_{\theta\theta}=R^2, g_{11}=\dfrac{\alpha'}{(2\pi)g_\ym^{2}}\,R^2 h(U)\) and \(B_{t\theta}=R\,h(U)\). We see that the proper size of the circle is small in the IR. For a translation-invariant background in \(\theta\), T-duality relates a Type IIB solution to a Type IIA solution via \cite{Buscher:1987sk}:
\begin{align}
&\tilde g_{\theta\theta}=\frac{1}{g_{\theta\theta}},\quad
\tilde g_{\theta\mu}=\frac{B_{\theta\mu}}{g_{\theta\theta}},\quad
\tilde g_{\mu\nu}=g_{\mu\nu}-\frac{g_{\theta\mu}\,g_{\theta\nu}-B_{\theta\mu}\,B_{\theta\nu}}{g_{\theta\theta}},\nonumber\\
&\tilde B_{\theta\mu}=\frac{g_{\theta\mu}}{g_{\theta\theta}},\quad
\tilde B_{\mu\nu}=B_{\mu\nu}-\frac{g_{\theta\mu}\,B_{\theta\nu}-g_{\theta\nu}\,B_{\theta\mu}}{g_{\theta\theta}},\quad
e^{\tilde\phi}=e^{\phi}\,g_{\theta\theta}^{-1/2}. \label{Buscher}
\end{align}
Apply \eqref{Buscher} to \eqref{F1metric}--\eqref{F1fields}, and
introduce the dual
coordinate \(\tilde x_1=\tilde R\,\tilde\theta\) with \(\tilde R=\alpha'/R\).
Then the type IIA metric becomes
\begin{align}
\frac{\d s^2}{\alpha'}
&=\frac{1}{(2\pi)g_{\ym}^{2}}\Big(\d U^2+U^2 \d\Omega_7^2\Big)
-\frac{1}{(2\pi)g_{\ym}^{4}}\;\frac{U^{6}}{d_1\,N}\;\d t^2\\
&\quad+\;(2\pi)\,d_1\,(g_{\ym}^{2})^{2}\,\frac{N}{U^{6}}\,
\Bigg(\d\tilde x_1-\frac{U^{6}}{\alpha'\,d_1\,g_\ym^{2}\,N}\,\d t\Bigg)^{\!2}\!,\\
\tilde B &= 0,\qquad e^{\tilde\phi}=\frac{1}{\sqrt{2\pi}\,g_\ym \,R}\,.
\end{align}
The dual radius is \(\tilde R=\alpha'/R\).
Somewhat surprisingly, we have found a pure-metric solution which carries momentum in the $\tilde{x}_1$ direction.
\end{exercise}

\begin{figure}
    \centering
\[ 
\begin{tikzpicture}
  \draw[thick]
    (-4.5,0.05) .. controls (1,0.1) and (1,2) ..
    (2,2) .. controls (3,2) and (3,0) .. (8.5,0.05);
  \begin{scope}[yscale=-1] %
  \draw[thick]
    (-4.5,0.05) .. controls (1,0.1) and (1,2) ..
    (2,2) .. controls (3,2) and (3,0) .. (8.5,0.05);
  \end{scope}
  \draw[fill=black] (2,0) circle (0.5);
  \draw[red,dashed,thick] (-4.5+2,-1) -- (-4.5+2,1);
  \draw[red,dashed,thick] (4.5+2,-1) -- (4.5+2,1);
  \draw[blue,dashed,thick] (2.5+2,-1) -- (2.5+2,1);
   \draw[blue,dashed,thick] (-2.5+2,-1) -- (-2.5+2,1);
  \node at (-0.5,-1.3) {$\frac{r}{\ell_p} \sim N^{1/9}$};
  \node at (-2.25,1.4) {$\frac{r}{\ell_p} \sim N^{1/7}$};
  \node at (5.5,1.1) {11d black};
  \node at (5.5,.6) {string};
  \node at (7.6,1.1) {IIA charged};
  \node at (7.6,0.6) {black hole};
  \node at (3.5,-0.25) {black hole};
  \node at (3.5,.25) {11d};
\end{tikzpicture}
\]
    \caption{A (not very) artistic depiction of the gravity dual at low energies  where $N \to \infty$, $\left(E/g^{2/3}_\ym\right) \to 0$ holding $ N\left(E/g^{2/3}_\ym\right)  = $ fixed and large. Deep in the IR is an 11d Schwarzschild-like black hole with radius $(r_\text{Schwarzschild}/\ell_p)^{16} \sim  N \left(E/g^{2/3}_\ym\right)  $. We have sketched the size of the M-theory circle. Far away $N^{1/7} \ll r /\ell_p \ll  N^{1/3}$, the dilaton is small and the appropriate description is the 10d Type IIA charged black hole. The upper limit $r/\ell_p \sim N^{1/3}$ is the 't Hooft scale --- larger radii represent the weakly coupled regime of the matrix model; the stringy corrections to the IIA description become appreciable. At the lower end of this limit, the dilaton/M-theory circle is $O(1)$ and we should lift to the black string solution in 11d. As we go further down the throat, we start to notice that the solution is localized along the M-theory circle. The inhomogeneity effects are important at $r/\ell_p \sim N^{1/9}$; once we are within a fixed number of Schwarzschild radii of the horizon, we can effectively ignore the compactification. }
    \label{fig:enter-label}
\end{figure}
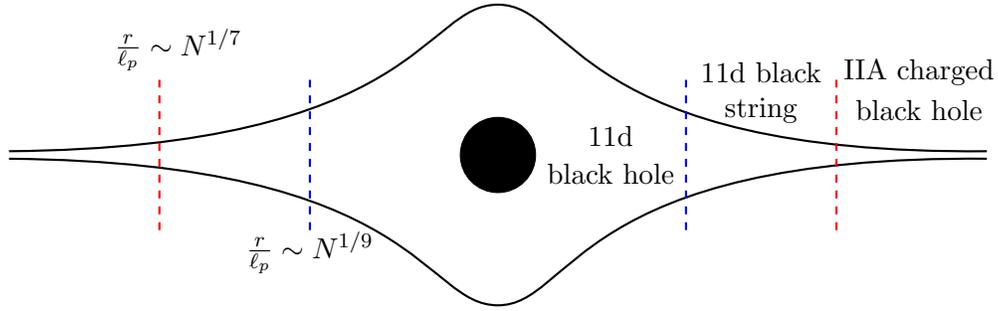

\subsection{Differentiate dictionary}
We now review a schematic derivation of the ``differentiate'' dictionary that is familiar in the AdS/CFT context (for a review in the AdS context, see section 4 of \cite{Klebanov:2000me}). 

Before taking the decoupling limit, we consider a background which, instead of being asymptotically flat, contains some supergravity waves. For simplicity, we could consider some (weak-field) dilaton waves incident on the D0-branes. Then we repeat the decoupling logic. From the gravity point of view, we are left with a perturbation of flat space $\otimes$ the throat-like region given by \eqref{gzz} but now with a non-trivial dilaton profile. The profile in the throat can be thought of as arising from non-trivial boundary conditions of the dilaton, which is an imprint from the joining of the throat to flat space.

On the boundary side, the (schematic) non-Abelian generalization of the  DBI action contains a coupling $e^{-\phi({t,X})} F^2$ where $F^2$ is the Yang-Mills field strength and $X^I$ are the transverse coordinates of the D$0$-brane\footnote{This is schematic because we are ignoring the fermions, and we are also being a bit sloppy about the precise non-Abelian generalization of the DBI action.}:
\begin{align}
    S_\text{SYM} \sim \int \d t \, \Tr (e^{-\phi(t,X^I(t))} \sqrt{- g(X^I(t))}  - A_0(t,X^I(t)) )
\end{align}
These scalars transform under the global $R$-symmetry of the SYM as SO($9$) fundamentals. The coupling to the dilaton wave induces a perturbation which leads to a deformation of the worldvolume action\cite{Klebanov:1997kc, Klebanov:1999xv} by terms schematically of the form: %
\begin{align}
    S_\text{SYM} \to   S_\text{SYM} + \mathcal{N} \sum_k \frac{1}{k!}\int \d t \,  \pd_{I_1} \cdots \pd_{I_k} \varphi \Tr (F^2 X^{(I_1} \cdots X^{I_k)}).
\end{align}
Here the indices $I_1\cdots I_k$ are symmetrized with traces removed\footnote{The indices should be symmetrized since the order of the derivatives acting on $\phi$ does not matter.}. The precise form of the operator (including the fermionic terms) can be obtained by acting with 4 supercharges on the operator 
\begin{align} \label{dbi_analysis}
    \mathcal{O}_{k+2} = \Tr X^{(I_1} \cdots X^{I_{k+2})},
\end{align}
where the parentheses on the indices indicate that the operator is in the traceless symmetric representation of SO(9).

Reasoning in this way, we learn that deforming the SYM theory by adding various terms to the Lagrangian is equivalent to changing the boundary conditions of various supergravity modes in the throat-like region. By differentiating with respect to a source term for these operators, we can compute correlation functions in the boundary theory using gravity. {\it A priori}, the gravity computation in this background seems complicated, but in fact we will see in the next section that the gravity problem is closely related to AdS. This implies for instance that 2-pt functions of the operator \eqref{dbi_analysis} will have simple power law correlation functions in the low temperature/large Euclidean-time Type IIA regime.

\subsection{Relation to AdS}
By introducing the radial coordinate $z=R_\ads \rho^{-5/2}$ (do not confuse this $z$ with the previous M-theory $z$), we may write the solution in an instructive form:
\begin{align} 
    \label{gzz}  \frac{ \d s^2}{\alpha'} &= \left(\frac{z}{R_\ads }\right)^{\frac{3}{5}} \left[R_\ads^2 \left(\frac{h(z)\, \d \tau^2+ h^{-1}(z) \d z^2}{z^2}\right)+\d \Omega_{8}^2\right], \\
    h&=1-\frac{z^d}{z_0^d}, \quad  d = 1+\frac{9}{5}, \quad 
    R_\ads = \frac{2}{5},  \\
    e^{-2\phi} &=  \left(\frac{d_0 N }{(2\pi)^2} \right)^{2} \left(\frac{z}{R_\ads}\right)^{\frac{-21}{5} },
   \quad  A_{0 } =  \frac{\ell_s d_0 N }{(2\pi)^2} \left(\frac{z}{R_\ads}\right)^{-\frac{14}{5}}.  \label{phi_sol}
\end{align}
It is instructive to consider the action associated with small fluctuations of the dilaton, e.g., $\phi = \phi_\text{s} + \varphi$, where the classical solution $\phi_\text{s}$ is given by \nref{phi_sol}. (The treatment below is somewhat heuristic\footnote{Among other concerns, we have not argued that $\varphi$ does not mix with other modes.}; for a detailed treatment, see \cite{Sekino:1999av} and also \cite{Biggs:2023sqw}.) Then expanding to quadratic order in $\varphi$, \eqref{actionII} gives %
\begin{align}%
   I \sim  \frac{4}{(2\pi)^7 (\alpha')^4} \int \d^{10} x \sqrt{g} e^{-2\phi_\text{s}}  (\nabla \varphi)^2.\label{probefield}
\end{align}
Decomposing $\varphi$ into scalar spherical harmonics $Y_k$ satisfying $\nabla^2_{S^n}Y_k=-k(k+n-1)Y_k$, we have:
\begin{align}\label{action}
    I&\propto \int \d^{8} \Omega \,  \d z \, \d \tau  \,  z^{1-d}  \left[ R^2_\ads \left( h (\pd_z \varphi)^2 +  \frac{(\pd_\tau \varphi)^2}{h} +(\vec{\nabla} \varphi)^2 \right) + k (k + 7) \varphi^2 \right] \\
    &=  \int \d^{8-p} \Omega \, \d^{d-1} \vec{x} \, \d z \, \d \tau\, \sqrt{g_\text{AdS}}  \left[ (\nabla_{\ads} \varphi)^2 +m_k^2 \varphi^2\right]  %
\end{align}
We see that these modes behave like massive fields in the AdS$_{d+1}$ black brane \cite{Kanitscheider:2009as, Biggs:2023sqw}, with an effective mass $m_k^2 = k(k+7)$. Note that we have introduced an additional $d-1$ spatial dimensions to account for the factors of $z$. (We implicitly assume that these extra dimensions are toroidal $T^{d-1}$ and that all fields are homogeneous on $T^{d-1}$, see \cite{Biggs:2023sqw}.)
One can then compute the scaling dimensions of these fields using the usual relation 
\begin{align}
    m^2 R^2_\ads = \Delta ( \Delta- d) \implies    \Delta = R_\ads (k + 2) + 2 \label{dim1}
\end{align}
Note here the scaling dimension is defined so that (in the zero-temperature limit) %
\begin{align} \label{2pt}
    \ev{\mathcal{O_\phi}(x) \mathcal{O_\phi}(0)} \sim \frac{1}{|x|^{2\Delta - (d-1) }}.
\end{align}
Here the additional factor of $d-1$ arises since $\mathcal{O}$ is a zero mode in the ``extra'' AdS dimensions.

\begin{boxcomment}{Scaling symmetry}
The solution \eqref{gzz}-\eqref{phi_sol} has an important scaling symmetry \cite{Kanitscheider:2009as, Biggs:2023sqw}:
\begin{align}
\tau \rightarrow \gamma \tau, \quad z \rightarrow \gamma z,
\end{align}
We keep $N$ and $g^2_\ym$ fixed under the scaling. 
Under this transformation, $g, \phi$ and $A_0$ change in such a way that the Type IIA action is {\it not} invariant. However, the action changes in a simple way:
\begin{align}
I \rightarrow \gamma^{-9/5} I.
\end{align}
This scaling ``symmetry'' is responsible for the simple dependence of the free energy on temperature. We can also use this scaling symmetry to characterize some perturbations to the solution. In particular, consider more general boundary conditions on some supergravity field $\chi \sim \chi_r(\tau) z^{\Delta}$ as $z \to 0$. Then $\chi_r \to \chi_r \gamma^{\Delta}$. On the other hand, the change in the gravity action (in the free field approximation) is
\begin{align}
I \sim \int \d \tau \, \d \tau' \chi_r(\tau) \chi_r(\tau') G(\tau-\tau').
\end{align}
According to the differentiate dictionary, $G$ is the boundary 2-pt function. Then demanding $I \to \gamma^{-9/5} I = \gamma^{-(d-1)} I $ we conclude that $G \to G \gamma^{-2 \Delta-(d-1)}$ or
\begin{align}
G \sim |\tau|^{-2\Delta-(d-1)}.
\end{align}
Please note that this argument does {\it not} work for massive fields, e.g., ones corresponding to massive stringy states, see \cite{Biggs:2023sqw} for more.
For recent work using this scaling symmetry, see \cite{Batra:2025ivy, Biggs:2025qfh, Bobev:2025idz}.
\end{boxcomment}

This gravity analysis shows there should be boundary operators that have conformal 2-pt functions \cite{Sekino:1999av, Kanitscheider:2008kd, Kanitscheider:2009as}. These operators should be BPS operators.
Let's recall the supersymmetric properties of the operator $\mathcal{O}_k$ that we obtained from the DBI analysis \eqref{dbi_analysis}. In the above $\mathcal{O}_k$ denotes any one of a number of operators that spans the symmetric traceless irrep of SO($9$). To analyze the action of the supercharge on this operator, it is convenient to pick one particular element of the irrep. We define the complex matrix 
\begin{align}
    \mathcal{Z}=X^{1} + \i X^{2},\quad O_\text{BPS} =  \Tr \mathcal{Z}^k.
\end{align}
Clearly $\mathcal{Z}^{\otimes k}$ transforms under SO($9-p$) as a symmetric tensor of rank $k$; it is also traceless since $\mathcal{Z} = \vec{y} \cdot \vec{X}$ and $\vec{y}$ is a null vector $\vec{y} \cdot \vec{y} = 0$.
Note that $[Q_\alpha, \mathcal{Z}] = (\gamma^1 + \i \gamma^2)_{\alpha \beta} \psi^\beta$. The matrix $(\gamma^1 + \i \gamma^2)$ has a $\mathcal{N}/2$ dimensional null space. (Here $\mathcal{N}$ is the dimension of the spinor irrep of SO($9-p$), e.g., $\mathcal{N}= 16$ for the D0 brane theory and famously $\mathcal{N} = 4$ for $p=3$.)
So any operator made out of $\mathcal{Z}$ is 1/2-BPS.
Since acting with a supercharge changes the dimension of an operator by $1/2$, together with \eqref{dim1}, this implies that $\Tr \mathcal{Z}^k$ or any traceless symmetric operator has dimension
\begin{align}
\label{eq:dim}
\Delta_{\mathcal{O}_k} = \frac{2}{5} k, \quad  \mathcal{O}_k = \Tr X^{(I_1} \cdots X^{I_{k})}.
\end{align}
Here we have used that $R_\text{AdS} = 2/5.$ This is the analog of the famous fact that in $\mathcal{N} = 4 $ SYM the scaling dimensions of similar operators $\Tr \mathcal{Z}^J$ are given by $\Delta = J$.

This relation to AdS is useful beyond just computing the correlation functions at zero temperature\footnote{More precisely, by zero temperature we mean a regime where the temperature is arbitrarily low in 't Hooft units but is not suppressed by powers of $N$ so that we are within the Type IIA regime. Similarly, Euclidean time separations are large in 't Hooft units but do not scale with $N$.}; in particular the relation to the AdS black brane is useful for computing the thermal 2-pt function (and in particular the quasi-normal modes, see \cite{Biggs:2023sqw}).

\subsection{Black hole thermodynamics \label{therm}}
Let us review what is known about the corrections to the black hole thermodynamics beyond pure supergravity. We expect the relevant corrections to the Type IIA effective action to have the form
\begin{align}
    -I \supset \frac{1}{(2\pi)^7 \ell_s^8} (\alpha')^3 \int \sqrt{g}\left[\# e^{-2 \phi}\left(R^4+\cdots\right)+\tilde{\gamma} \left(R^4 +\cdots \right)\right]
\end{align}
Here $R^4+\cdots$ is meant to indicate a non-linear correction that involves 8 derivatives. Note that it is insufficient to consider just the 8-derivative terms involving the metric; for this background we in principle need to know the full non-linear 8-derivative correction that involves the RR 1-form $A_\mu$, e.g., terms $\sim F^8$ which are at present unknown.

On the other hand, the 1-loop term $\sim \tilde\gamma$ (which also involves the Ramond-Ramond field, etc.) arises from dimensional reduction of the 8-derivative pure metric term in M-theory \cite{Green:1997as}:
\begin{align} \label{oneloopmtheory}
-I= & \frac{1}{(2\pi)^8 \ell_P^9 } \int \sqrt{g} R+\frac{\gamma}{l_p^3} \int \sqrt{g}  \left(t_8 t_8 R^4 + \frac{1}{4} E_8+\cdots\right)\\
& \gamma=\frac{1}{2^{19} \times 3^2 \times \pi^6}
\end{align}
This is a tidy way to specify the 1-loop term in the Type IIA effective action because from the 11d point of view, the solution is pure metric. 

Now let us discuss the implications of this for the black hole thermodynamics. To first order, we simply evaluate the effective action on the solution. To evaluate the 1-loop term, one can simply use the 11d black string solution \eqref{blackString}, see \cite{Hyakutake:2013vwa, Hyakutake:2014maa} and the Appendix of \cite{Biggs:2023sqw}. Although we do not know the full tree-level correction, knowing that the leading correction $\sim (\alpha')^3$ together with dimensional analysis allows us to estimate the temperature dependence of the correction.

To summarize the results, it is convenient to measure energy and temperature in 't Hooft units $\tilde{E} = E/(g_\ym^2 N)^{1/3},  \tilde{T} = T / (g_\ym^2 N)^{1/3}$. Then we have a double expansion in low temperature and large $N$:
\begin{align}\label{et-thermo}
    \frac{\tilde{E}}{N^2}=\frac{\left(a_0 \tilde{T}^{\frac{14}{5}}+a_1 \tilde{T}^{\frac{23}{5}}+a_2 \tilde{T}^{\frac{29}{5}}+\cdots\right)}{N^0}+\frac{\left(b_0 \tilde{T}^{\frac{2}{5}}+b_1 \tilde{T}^{\frac{11}{5}}+\cdots\right)}{N^2}+\mathcal{O}\left(N^{-4}\right) 
\end{align}
Only $a_0$ and $b_0$ are known analytically. In exercise \ref{thermodynamicsEx}, you were asked to show that $a_0 =  \frac{9}{14} 4^{13 / 5} 15^{2 / 5}(\pi / 7)^{14 / 5} $. %
To understand the exponent $23/5$ associated with $a_1$, note that the curvature at the horizon scales $\alpha' R \propto \tilde{T}^{3/5}$, so a term like $\sim R^4$ will give an additional factor of $T^{9/5}$ relative to the Einstein-Hilbert term $\sim R$. Furthermore, the sub-leading $a_2$ comes from a 12-derivative term. So the subleading exponent associated with $a_2$ is given by
\begin{align}
\frac{12-8}{8-2} \times \frac{23-14}{5} + \frac{23}{5} = \frac{29}{5}.
\end{align}
The $b_0$ term can be obtained by integrating the $R^4$ correction to M-theory on the background \eqref{blackString}, see  \cite{Hyakutake:2013vwa, Hyakutake:2014maa} and the Appendix of \cite{Biggs:2023sqw}. The $b_1$ term comes from $\sim D^6 R^4$, which again leads to a $T^{9/5}$ suppression relative to the $R^4$ M-theory term.

\subsection{Monte Carlo}
An extremely important tool for studying this model is lattice Monte Carlo \cite{Anagnostopoulos:2007fw, Hanada:2008ez, Catterall:2008yz, Catterall:2009xn, Filev:2015hia, Rinaldi:2021jbg, Pateloudis:2022ijr, Bodendorfer:2024egw}. Most of these studies attempt to approach the strongly coupled 't Hooft regime by putting the theory on a Euclidean circle and working in the path integral formalism. To do so, one discretizes Euclidean time, and then integrates out the fermions to derive a measure for the bosonic matrices and the gauge field. 

There are two important subtleties when doing these simulations. First, after integrating out the fermions there is no guarantee that the resulting measure is positive, i.e., that it can be interpreted as a probability measure that can then be efficiently sampled from. In practice, one simply samples from the absolute value of the measure\footnote{This is sometimes referred to as the ``quenched'' approximation in the literature.}; {\it a priori} this seems like a dramatic modification to the problem; however, various authors \cite{Catterall:2008yz, Catterall:2009xn, Filev:2015hia} give evidence that the phase of the measure is sharply peaked at some values of temperature and $N$, although it is unclear whether fluctuations in the phase are parametrically suppressed in the strongly coupled 't Hooft regime.
Second, since the black hole phase is only metastable at finite $N$, one must regulate the problem by effectively putting the system in a box. In the modern approach, one does this by turning on the BMN mass deformation term which lifts all the flat directions \cite{Berenstein:2002jq}. 

There is also a Monte Carlo prediction \cite{Pateloudis:2022ijr} for $a_1 = -9.90 \pm 0.31$ and even $a_2 = 5.78 \pm 0.38$, see also \cite{Berkowitz:2016jlq}. As we have already discussed, just predicting $a_1$ using string theory is an interesting challenge; in principle, it could be extracted from a tree-level 8-pt amplitude, but one could also hope for a less cumbersome method. Impressively, \cite{Pateloudis:2022ijr} also claim to reproduce the 1-loop value of $b_1$, which is the first hint of 11d M-theory physics. See box \eqref{mtheory_approach} for some discussion on other approaches to computing these M-theory corrections.

\section{The matrix model}
The D0 brane matrix quantum mechanics consists of 9 bosonic matrices $X_I$ and 16 fermionic matrices $\psi_\alpha$, which transform under an SO(9) $R$-symmetry in the fundamental and spinor representations. The matrix model was introduced and studied before the BFSS conjecture \cite{Claudson:1984th, deWit:1988wri, Douglas_1997}.  All matrices are taken to be Hermitian and traceless; they satisfy the canonical commutation relations:
\begin{equation}
\begin{split}
\label{canonical}
 \quad [X^I_{ij},P^J_{kl}] = \i \, \delta_{il} \delta_{jk} \delta^{IJ} , \quad \{  \psi_{\alpha,ij} , \psi_{\beta,kl} \} =   \delta_{\alpha \beta}  \delta_{il} \delta_{jk}   %
\end{split}
\end{equation}
In this review, there is no distinction between upper and lower indices.
The Hamiltonian is
\begin{align}\la{ham}
H &= \frac12  \operatorname{Tr}  \left(g^2_\ym   {P_I^2}-\frac{1}{2g^2_\ym}\left[X_I, X_J\right]^2-\psi_\alpha \gamma^I_{\alpha \beta} \left[ X_I,\psi_\beta\right] \right) .
\end{align}
In the above expression, there is an implicit sum over $I,J$. 
With these conventions, $X$ has units of energy and $g^2_\ym$ has units of $E^{3}$.
We can take the SO(9) gamma matrices $\gamma^I$ to be real, traceless, and symmetric. 
To compare with the BFSS literature, it is convenient to perform the canonical transformation $X = \tilde{X}/(2\pi \ell_s^2)  , P = (2\pi \ell_s^2) \tilde{P} $ so that $\tilde{X}$ has units of length and
\begin{align}
    H &= \frac12  \operatorname{Tr}  \left(g_s \ell_s  {\tilde{P}_I^2}-\frac{1}{2(2\pi)^2 g_s \ell_s^5 }\left[\tilde{X}_I, \tilde{X}_J\right]^2- \frac{1}{2\pi \ell_s^2} \psi_\alpha \gamma^I_{\alpha \beta} \left[ \tilde{X}_I,\psi_\beta\right] \right), \\
    &= \frac{R}{2}  \operatorname{Tr}  \left(  {\tilde{P}_I^2}-\frac{1}{2 (2\pi \ell_p^3)^2  }\left[\tilde{X}_I, \tilde{X}_J\right]^2- \frac{1}{2\pi \ell_p^3}  \psi_\alpha \gamma^I_{\alpha \beta} \left[ \tilde{X}_I,\psi_\beta\right] \right) . \label{planckUnits}
\end{align}
Here we take the matrices to be traceless and Hermitian so that they transform under the gauge group SU($N$). We could also include the zero mode, which would just be a free non-relativistic particle\footnote{In these units the U(1) mode is governed by $H = \frac{R}{2 N}p^2$. If we fix $p^2 \ell_p^2 $, then $H \ell_p = \frac{R}{2 \ell_p N} (p^2 \ell_p^2)$}. This model has 16 supercharges which transform as spinors under the SO(9) global symmetry.
The 16 Hermitian supercharges are
\begin{align} \label{eq:susy}
    Q_\alpha &= g_\ym \Tr P_I \gamma^{I}_{\alpha \beta} \psi_\beta - \frac{\i}{2g_\ym} \Tr [X^I, X^J]\gamma^{IJ}_{\alpha \beta}\psi_\beta,
\end{align}
They satisfy the supersymmetry algebra
\begin{align}\label{eq:gaugegen}
    \{Q_\alpha, Q_\beta\} &= 2 \delta_{\alpha \beta} H +2 \gamma^I_{\alpha \beta} \Tr X^I C\\
    C &= -\i [X^I, P^I]-\psi_{\alpha} \psi_{\alpha} - N {\bf 1}
\end{align}
In the above equation, $C_{ij}$ is a generator of SU$(N)$ symmetry, where each matrix transforms in the adjoint representation. By choosing the matrices to transform in SU$(N)$ as opposed to U($N$), we have removed the center of mass degree of freedom.
We have the option to gauge or ungauge the model. For a quantum mechanical system this just means whether we take the Hilbert space to be the SU$(N)$ invariant sector or whether we include arbitrary states that transform non-trivially under SU($N$). Note that $C \ne 0 $ violates SUSY. For more, see \cite{Maldacena:2018vsr}.

\begin{exercise}{Dimensional reduction}
Dimensionally reduce flat space 10D SYM, $\mathcal{N}=1$ SYM or 4D $\mathcal{N}=4$ SYM and check that it gives BFSS. Consider $\mathcal{N}=4$ SYM on $S^3 \times \mathbb{R}$. The theory on $S^3$ has an SO(4)$\simeq \mathrm{SU}(2)_L \times \mathrm{SU}(2)_R$ symmetry. Show that at the classical level, truncating to the modes that preserve $\mathrm{SU}(2)_R$ leads to the massive Berenstein-Maldacena-Nastase (BMN) matrix model \cite{Berenstein:2002jq}; see \cite{Kim:2003rza}. 
\end{exercise}

\subsection{The spectrum at finite \texorpdfstring{$N$}{N} \label{spectrumN}}
Here is what is believed to be true about the model at finite $N$:
\begin{enumerate}
    \item There exists a unique, normalizable zero-energy ground state that preserves $Q_\alpha$ and is rotationally invariant. The evidence for this is a rather subtle index computation, see \cite{Yi:1997eg, Sethi:1997pa} for $N=2$ and \cite{Moore:1998et, Konechny:1998vc} for results for more general $N$. Furthermore, \cite{Sethi:2000zf} argued that all ground states must be SO(9) singlets. This implies that all ground states are bosonic, $(-1)^F = 1$, and, combined with the index result, implies that there is a unique normalizable zero-energy state for $N=2$.
    Naively, one would think that the Witten index $\Tr (-1)^F e^{-\beta H}$ as $\beta \to 0$ can be computed using the IKKT matrix integral \cite{Moore:1998et}, but this is not quite right due to the flat directions.
    \item There are power law tails of the wave function; these can be studied in a $1/r$ expansion. These tails are associated with splitting the $N \times N$ matrix into approximately block diagonal configurations, with $N_1 \times N_1$ and $N_2 \times N_2$ sub-matrices in the ground state \cite{Plefka:1997xq,Frohlich:1999zf,Hoppe:2000tj, Hasler:2002wt,Lin:2014wka}. Splitting into two sub-matrices is associated with a $1 / r^9$ power law tail, where $r = |x_1-x_2|$ is the relative separation between the blocks. The ground state wavefunction enjoys a further ``factorization'' property where each block hierarchically splits into smaller sub-blocks \cite{Hasler:2002wt, Lin:2014wka}.
    \item All other states are scattering states, with the continuum starting at $E>0$. There are no bound states with energy $E >0$ (although at large $N$ it is believed that there are metastable states, see Exercise \ref{blackholebox}). The in/out scattering states have the following block diagonal form:
\begin{align}
\renewcommand{\arraystretch}{2} %
\begin{pmatrix}
\underbrace{\psi(\vec{X}_1, \vec{x}_1)}_{N_1\times N_1} & 0                          & 0                          \\[2ex]
0                          & \underbrace{\psi(\vec{X}_2, \vec{x}_2)}_{N_2\times N_2} & 0                          \\[2ex]
0                          & 0                          & \underbrace{\psi(\vec{X}_3, \vec{x}_3)}_{N_3\times N_3}
\end{pmatrix}, \quad \psi(\vec{X}_i , \vec{x}_i)  = \psi_{0}(\vec{X}_i) e^{\i \vec{p_i} \cdot \vec{x}_i}\label{blockScatter}
\end{align}
Here $\psi_0$ is the ground state wavefunction of a smaller BFSS matrix model with $N = N_i$. $X_i$ is an $N_i \times N_i$ traceless Hermitian matrix, and $x_i$ is the trace of the block, e.g., the center of mass. The zeros indicate that the wavefunction is peaked around zero (in some gauge), e.g. that the overall wavefunction is peaked around matrices that are simultaneously block-diagonalizable. We have depicted a situation where there are 3 blocks, but in general there could be any number of blocks $2 \le n_\text{blocks} \le N $.

We have also suppressed the fermions in the above notation; the ground state wavefunction should really be written $\psi_0(X^I_i, \psi^\alpha_i)$ where $\psi^\alpha_i$ are all traceless matrices. Each center-of-mass coordinate $x_i$ comes with 16 fermionic superpartners $\psi_\alpha$. Quantizing them gives a Hilbert space of dimension
\begin{align} \label{polarizationStates}
    \dim{\mathcal{H}} = 2^{16/2} = 256 = 44 +84 + 128.
\end{align}
To fully specify the asymptotic state, we should also specify the state in this Hilbert space. We have written the decomposition of 256 in anticipation of the BFSS conjecture \eqref{bfss_conj}.

Since each block is in its internal ground state, the total energy of the scattering state comes only from center-of-mass motion:
\begin{align}
    E = \sum_i \frac{R}{2N_i} \vec{p}_i^2
\end{align}
and is independent of the polarization state.
\end{enumerate}

\section{Scattering and the BFSS conjecture}

We would like to understand the gravity dual of the scattering setup described in the previous section.

\subsection{Scattering in the 't Hooft limit}
A generalization of the extremal solution \eqref{string_frame} is to multi-center solutions:
\begin{align}
\d s^2 &= - H^{-1/2}(r)   \d t^2 + H^{1/2}(r) ( \d r^2 + r^2 \d \Omega_8^2),\\
     e^{2 \phi} &= g_\mathrm{s}^{2} H^{3/2}, \\
H(\vec{r}) &=1+\sum_{i=1}^k \frac{d_0 g_s N_i  \ell_s^7}{(2\pi)^2\left|\vec{r}-\vec{r}_i\right|^{7}}, %
\end{align}
The BPS condition $Q=M$ implies that the gravitational force cancels the Coulomb force between the black holes and therefore one can place these black holes anywhere with respect to each other.

By repeating the decoupling arguments, we conclude that this configuration is represented by a state in the matrix quantum mechanics where the $X_i$ have vevs given by $r_i$. The charge of each black hole is represented by $N_i$, the size of each block diagonal matrix.

Since the multi-center solutions exist for any choice of $r_i$, it is natural to consider a generalization of these solutions where the black holes are moving slowly with respect to each other. In particular, by the decoupling logic, these configurations would be relevant for the scattering problem in the matrix quantum mechanics.
There is a general procedure for generalizing these multi-center solutions to slowly moving solutions \cite{Ferrell:1987gf}. We promote $\vec{r}_i \to \vec{r}_i(t) $. Then in the small velocity expansion, 
\begin{align}
        \d s^2 &= - H^{-1/2}(\vec{r}(t))   \d t^2  + 2 N^i \d x^i \d t  + H^{1/2}(\vec{r}(t)) \d x^i \d x^i ,\\
        A &= A_0 \d t + A_i \d x^i
\end{align}
Here $A_i$ and $N^i$ are determined by solving the constraints. Actually this calculation was already done by Shiraishi \cite{Shiraishi:1992nq}; in the notation of \cite{Shiraishi:1992nq} the values relevant\footnote{The paper claims that $a=1$ for string theory, but this is a typo. In string frame RR fields should not be coupled to the dilaton.} for the D0 brane computation are $a^2 = N = 9$. In principle, 
\begin{align} \label{effPot}
    \mathcal{L} \sim V(r) + \hf g(r) v^2 +  v^4 F(r) + \cdots %
\end{align}
We have seen that the cancellation of electric and gravitational forces implies $V = 0$. Actually SUSY implies that the $g=1$, consistent with the results of \cite{Shiraishi:1992nq}. It would be interesting to compute $F(r)$ by going to higher orders in the low velocity expansion. (This is similar to a post-Newtonian approximation, except that we do not assume that the metric is close to flat.)

A slightly different regime is to consider scattering with a large cluster of D0's and a small number of ``probe'' D0's. In the 't Hooft limit, we can view this process as being governed by the probe D0 brane action in the extremal D0 black hole background \cite{Becker:1997xw}.
The DBI action for a probe D0 brane in this background is   
\begin{align}
	I &= -T_0 \int_{\mathrm{D} 0} (\d \tau  \, e^{-\phi} \sqrt{-g_{\mathrm{D}0}} - A_0 )\\
  &= -  \frac{N d_0}{ (2\pi)^2 }\ \int \d \tau \,  \rho^7  \left(\sqrt{1 - \frac{ v^2} {\rho^7} } -1\right), \quad v^2 = \dot{\rho}^2 + \rho^2 \cos^2 \theta \dot \phi^2 \label{SDBIrho} \\
  &\approx \frac{N d_0}{ (2\pi)^2 }\int \d \tau \, \left[ \frac{v^2}{2} + \frac{v^4}{8 \rho^7} + \frac{v^6}{16 \rho^{14}} \cdots \right] \label{v4expansion}
\end{align}
We see that we recover the above facts, namely that $V = 0, g=1$ and that $F \propto 1/\rho^7$. A trivial generalization of this computation is to replace $N \to N_1 N_2$ if we have multiple D0 branes in the gravity background with $N_1 \gg 1$. 

\begin{exercise}{Probe D0 brane scattering}
    Study the solutions\footnote{I thank Gauri Batra for discussions about these trajectories.} of \eqref{SDBIrho}. Show that there are  ``scattering'' trajectories in Lorentzian signature where the probe D0 brane approaches $\rho \to \pm \infty$ when $\tau \to \pm \infty$.
\end{exercise}

The result \eqref{v4expansion} can be reproduced from the BFSS matrix model. To compare with the literature, we should convert back to $r,t$ coordinates using $\rho = (r/\ell_s )/( 60 \pi^3  g_s N_1)^{1/3}  $ and $\tau = ( 60 \pi^3  g_s N_1)^{1/3} t/\ell_s $ %
\begin{align}
    I &= - \frac{N_2 R}{60\pi^3 N_1\ell_p^9}\int \d t\,r^7
    \left(\sqrt{1-\frac{60\pi^3N_1\ell_p^9\dot r^2}{R^2r^7}}-1\right)\\
    &\approx \int \d t\,\left(
    \frac{N_2}{2R}\dot r^2
    + \frac{15N_1N_2(2\pi)^3\ell_p^9\dot r^4}{16r^7R^3}
    + \frac{225}{64}\frac{N_1^2N_2\ell_p^{18}(2\pi)^6\dot r^6}{r^{14}R^5}
    + \cdots\right) \label{v4expansion2} %
\end{align}
The $v^4$ and $v^6$ terms have been matched (including the precise numerical coefficient\footnote{To compare with \cite{Banks:1996vh, Becker:1997xw} one should shift $\ell_p^3 \to \ell_p^3/(2\pi)$.}) to a 1-loop and 2-loop computation in the matrix theory \cite{Banks:1996vh, Becker:1997wh, Becker:1997xw}. 
The idea is that we separate the degrees of freedom into slow and fast modes and use a Born-Oppenheimer approximation. The fast modes are harmonic oscillators with $\omega = |X_1-X_2|/\alpha'$, which can be interpreted as arising from open strings connecting the separated D0 branes. For this calculation, it is easy to work in the Hamiltonian formalism (see \cite{Komatsu:2024vnb} for a recent discussion).

Let us comment on the loop counting parameter in this approach. We focus on two matrices to illustrate the approach. For small $w_x$ we have 
\begin{align}
    X_{2 \times 2}^1 = x \sigma_z + w \sigma_x \quad 
    X_{2 \times 2 }^I = X^I \sigma_z  ,\\
    [X,Y] \approx w X^I [\sigma_z, \sigma_x] \RA  \Tr [X,Y]^2 \approx w^2 X^I X^I
\end{align}
If the separation between matrices is large $X^I X^I$ is large, then we see that the $w$ mode is very heavy. The commutator square interaction gives us an oscillator term
\begin{align}
    \sim \frac{1}{2R} \dot{w}^2 + \frac{R w^2}{\ell_p^6} \tilde{X}^I \tilde{X}^I  %
\end{align}
This oscillator has a characteristic frequency given by the energy of a stretched string:
 \begin{align} \label{eq:oscfreq}
     \omega  \sim \frac{R }{\ell_p^3} r \sim \frac{1}{\alpha'} |X_1-X_2|,
 \end{align}
\begin{boxcomment}{The double expansion in loops and derivatives}
Following Becker, Becker, Polchinski, and Tseytlin
\cite{Becker:1997xw}, the two dimensionless parameters are
\begin{align}\label{loop}
 \lambda_\text{eff}\equiv\frac{d_0 g_\ym^2N_1}{U^3}
 =\frac{1}{\rho^3},
 \qquad
 a\equiv\frac{\dot U}{U^2}=\frac{\dot\rho}{\rho^2},
 \qquad
 \rho\equiv\frac{U}{(d_0g_\ym^2N_1)^{1/3}}.
\end{align}
Here \(\lambda_\text{eff}\) counts loops and an expansion in \(a^2\) is a derivative expansion:
\begin{align}
 \widehat{\mathcal L}_0
 &\sim c_{00}N_1N_2\dot\rho^2,\nonumber\\
 \widehat{\mathcal L}_1
 &\sim N_1N_2\dot\rho^2\lambda_\text{eff}\left(
 \cancel{c_{10}}+c_{11}a^2+c_{12}a^4+\cdots\right),\nonumber\\
 \widehat{\mathcal L}_2
 &\sim N_1N_2\dot\rho^2\lambda_\text{eff}^2\left(
 \cancel{c_{20}}+\cancel{c_{21}a^2}+c_{22}a^4+\cdots\right),\nonumber\\
 \widehat{\mathcal L}_3
 &\sim N_1N_2\dot\rho^2\lambda_\text{eff}^3\left(
 \cancel{c_{30}}+\cancel{c_{31}a^2}+\cancel{c_{32}a^4}
 +c_{33}a^6+\cdots\right).
\end{align}
Equivalently, the same expansion can be written entirely in terms of
\(\rho\) and \(\dot\rho\):
\begin{align} \label{columnSlash}
 \widehat{\mathcal L}_0
 &\sim c_{00}N_1N_2\dot\rho^2,\nonumber\\
 \widehat{\mathcal L}_1
 &\sim N_1N_2\left(
 \cancel{c_{10}}\frac{\dot\rho^2}{\rho^3}
 +c_{11}\frac{\dot\rho^4}{\rho^7}
 +c_{12}\frac{\dot\rho^6}{\rho^{11}}+\cdots\right),\nonumber\\
 \widehat{\mathcal L}_2
 &\sim {N_1N_2}\left(
 \cancel{c_{20}}\frac{\dot\rho^2}{\rho^6}
 +\cancel{c_{21}}\frac{\dot\rho^4}{\rho^{10}}
 +c_{22}\frac{\dot\rho^6}{\rho^{14}}+\cdots\right),\nonumber\\
 \widehat{\mathcal L}_3
 &\sim {N_1N_2}\left(
 \cancel{c_{30}}\frac{\dot\rho^2}{\rho^9}
 +\cancel{c_{31}}\frac{\dot\rho^4}{\rho^{13}}
 +\cancel{c_{32}}\frac{\dot\rho^6}{\rho^{17}}
 +c_{33}\frac{\dot\rho^8}{\rho^{21}}+\cdots\right).
\end{align}
In writing the above expansion, we have already omitted any static potential.
We have also crossed out terms which must vanish for the BFSS conjecture to hold.
For the 2-body effective action, supersymmetric nonrenormalization
theorems \cite{Paban:1998ea,Paban:1998qy} imply that the required cancellations do in fact occur for the first three columns in \eqref{columnSlash}.
\end{boxcomment}

For example, the 1-loop matrix theory computation which reproduces the $\dot{r}^4/r^7$ term in \eqref{v4expansion2} {\it a priori} should be corrected by a function $f_4(\lambda_\text{eff}) = 1 + c_1 \lambda_\text{eff} + \cdots $. The perturbative computation only guarantees that $f_4 = 1$ at small $\lambda_\text{eff}$ but to compare with Type IIA gravity, we need to know $f_4$ as $\lambda_\text{eff} \to \infty$, e.g., $N^{1/7} \ll  r/\ell_p \ll N^{1/3}$, see exercise \ref{thermodynamicsEx}. This corresponds to strong 't Hooft coupling $1 \ll \lambda_\text{eff} \ll N^{4/7} $. 
How are we supposed to extrapolate from weak coupling to strong coupling? Here one appeals to extended supersymmetry \cite{Paban:1998ea, Paban:1998qy}, which forbids certain corrections to the effective potential. This is enough to show that $V=0$ and $g =1$ in \eqref{effPot} and that $f_4$ is constant and can therefore be compared with the strong coupling (gravity) prediction. Similarly, the $\dot{r}^6$ term is protected and in fact the coefficient of the $\dot{r}^6$ term is determined\footnote{The $v^6$ terms are determined in the SU(2) effective theory which is relevant for $2 \to 2$ scattering. For higher-pt scattering amplitudes, one needs to consider larger gauge groups, where there are unfixed couplings \cite{Sethi:1999qv} for the $v^6$ terms but the $v^4$ terms remain protected \cite{Lowe:1998vu, Sethi:1999qv}. I thank Savdeep Sethi for discussions.} by lower derivative terms in the effective action \cite{Paban:1998qy}.

\subsection{Scattering beyond the 't Hooft limit }

In the previous section, we studied scattering at strong 't Hooft coupling. Perhaps even more interesting is to consider the ultra-strongly coupled limit 
\begin{align}
    r/\ell_p = \text{fixed}, \quad N \to \infty \label{ultraStrong}
\end{align} 
Here we are thinking of $r$ as the impact parameter $b$ in the scattering process. 
This is far beyond the conventional 't Hooft regime; it is more analogous to the limit $N\to\infty$,  $g^2_\ym = \text{fixed}$ in $\mathcal{N}=4$ SYM \cite{Susskind:1998vk, Polchinski:1999ry}. Indeed, the loop-counting parameter \eqref{loop} is then
$ \lambda_\text{eff}\sim N\left(\frac{\ell_p}{r}\right)^3\sim N.$
Based on the gravity picture, one should expect this to probe 11-d scattering. This is indeed what BFSS conjectured. From the modern point of view, their conjecture is actually even bolder, as they assume that the amplitude is not contaminated by any effects that occur in the weak coupling/stringy region.

Let us now state precisely the BFSS conjecture \cite{Banks:1996vh}:
\begin{align}
    \lim _{N_i \rightarrow \infty, R \rightarrow \infty} &\mathcal{A}_{\text{SYM}}\left(N_1, \vec{p}_1 ; \cdots ; N_n, \vec{p}_n\right)=(2 \pi R)^{1-\frac{n}{2}} \mathcal{A}_{\text {M-theory}}\left(p_1^\mu, \cdots, p_n^\mu \right), \label{bfss_conj} \\
    \ell_p (p_-)_i &= - \ell_p N_i/R = \text{fixed} \label{scalingBFSS} \\
    2 &p_+ p_- - \vec{p}^2 = 0
\end{align}
Here the RHS is the M-theory amplitude in asymptotically flat 11d spacetime.
For this conjecture to be true, it is necessary that the only stable particles in M-theory are massless. The fermionic zero modes encode their polarization states. In particular, we interpret the RHS of \eqref{polarizationStates} as follows: 44 is the dimension of rank 2 traceless symmetric tensors of SO(9) (the little group in 11d) and 84 is the dimension of rank 3 fully anti-symmetric tensors. 128 is a fermionic irrep of spin(9). These are precisely the states associated with the graviton, the 3-form gauge field $A_{\mu \nu \rho}$, and the gravitinos in M-theory.

The conjecture is sometimes stated as involving the limit $R\to\infty$, but we prefer to state it in terms of dimensionless quantities. Indeed, the dimensionless energy according to \eqref{scalingBFSS} and Table \ref{table1} is
\begin{align} \label{ultraLow}
 \frac{H}{(g_\ym^2)^{1/3}}
 \sim(p_+\ell_p)\frac{\ell_p}{R}\sim\frac{1}{N}.
\end{align}
This is precisely the energy at which an eleven-dimensional Schwarzschild black hole of fixed
radius in Planck units can appear as an intermediate state
\cite{Itzhaki:1998dd} in the scattering process.  From the holographic description of the localized
eleven-dimensional black hole, denoting its Schwarzschild radius and mass by $r_{\rm S}$ and $M_{\rm S}$, its energy above extremality
scales as
\begin{align}
 \frac{\Delta E}{(g_\ym^2)^{1/3}}
 &\sim \frac{1}{N}\left(\frac{r_{\rm S}}{\ell_p}\right)^{16}
 \sim \frac{(M_{\rm S}\ell_p)^2}{N},
 \qquad M_{\rm S}\ell_p\sim\left(\frac{r_{\rm S}}{\ell_p}\right)^8 .
\end{align}
It might seem peculiar that $\Delta E \propto M_{\rm S}^2$, but recall that for a boosted black hole with $N$ units of KK-momentum on a circle of radius $R$,
\begin{align}
 \Delta E
 &\simeq\sqrt{\left(\frac{N}{R}\right)^2+M_{\rm S}^2}-\frac{N}{R}
 \simeq\frac{R M_{\rm S}^2}{2N}+\cdots .
\end{align}
This reproduces the relation
$R/\ell_p^2=(4\pi^2g_\ym^2)^{1/3}$, see Table \ref{table1}.  
For a black hole with mass fixed in Planck units $M_{\rm S} \propto 1/\ell_p$, the dimensionless energy $\Delta E/(g_\ym^2)^{1/3} \sim {1/N}$, which is consistent with \eqref{ultraLow}.
For comparison, in the 't Hooft regime with the
dimensionless velocity $v^2$ in \eqref{v4expansion} fixed,
\begin{align}
 \frac{H}{(g_\ym^2N_1)^{1/3}}\sim N_1N_2v^2.
\end{align}
Thus the BFSS energies are parametrically smaller than the scattering
energies in the 't Hooft regime. 

\begin{exercise}{Gravitational scattering}
    Recall from Alexander Zhiboedov's lectures that in $d>4$, Einstein gravity in the limit $s = \text{fixed}, t \to 0$ gives the amplitude
    \begin{align}
 \mathcal{A}_\text{tree}(s,t) \sim - \frac{8 \pi G_N s^2}{t} 
    \end{align}
    Here $16 \pi G_N = (2\pi)^8 \ell_p^9$.
    Work out the 11d kinematics and show that $s= - N_1 N_2 \dot{r}^2/R^2  , t =- \vec{k}^2$. Then
    \begin{align}
        V  &= \frac{1}{2\pi R} \prod_{i=1}^{4} \frac{1}{\sqrt{2 E_i}}\int \frac{\d^9 k }{(2\pi)^9 } e^{\i \vec{k} \cdot \vec{r}} \mathcal{A}_\text{tree} \\
        &= \frac{15 N_1 N_2 (2\pi)^3 \ell_p^9 \dot{r}^4}{16 R^3 r^7}
    \end{align}
    where you may use 
    $\int \frac{\d^9 \vec{k}}{(2 \pi)^9} e^{\i \vec{k} \cdot \vec{r}}|\vec{k}|^{-2}=\frac{15}{2(2 \pi)^4} r^{-7}$. This agrees with the second term in \eqref{v4expansion2}. For help and/or the generalization to the $v^6/r^{14}$ term see \cite{Becker:1997cp}.
\end{exercise}

\begin{boxcomment}{Other approaches to M-theory. \label{mtheory_approach}}
We can also take the flat space limit of AdS/CFT \cite{Polchinski:1999ry, Susskind:1998vk, Penedones:2010ue} and extract flat space amplitudes. Note that this also involves a large $N$, ultra-strongly coupled limit.
\begin{enumerate}
    \item M2 branes: the low energy limit of M2 branes is given by a 3D superconformal field theory, the ABJM theory \cite{Aharony:2008ug}. This theory is dual to AdS$_4 \times S_7/\mathbb{Z}_k$. Using numerical conformal bootstrap (with the help of inputs from localization), one can compute stress-energy tensor correlation functions and study the flat space limit \cite{Chester:2018aca, Binder:2018yvd, Chester:2024bij}. The numerical CFT bootstrap seems promising for computing even non-protected terms to the M-theory effective action.
    \item M5 branes: the low energy limit of M5 branes is governed by the 6D (2,0) theory. The holographic dual is AdS$_7 \times S_4$ \cite{Maldacena:1997re}. The conformal anomaly\footnote{The $A_{N-1}$ theory has central charge $c(A_{N-1}) = 4N^3 -3N-1$. The $-3N$ term should be reproducible from the bulk $R^4$ term, according to \cite{Tseytlin:2000sf}.} of the 6D CFT at finite $N$ has been computed \cite{Intriligator:2000eq, Beem:2014kka, Ohmori:2014kda, Cordova:2015vwa, Cordova:2016cmu} and the sub-leading $N$ dependence should be reproduced by the $R^4$ term in M-theory \eqref{oneloopmtheory}, see \cite{Tseytlin:2000sf}. The $R^4$ term was carefully reproduced in \cite{Chester:2018dga} by considering 4-pt functions of 1/2-BPS operators. See also \cite{Beem:2015aoa, Alday:2020tgi} for bootstrapping this theory.
    \item Although this has never been worked out in detail, one could in principle extract the M-theory S-matrix from correlation functions of the supergravity modes in BFSS using a similar logic to the AdS/CFT flat space limit \cite{Polchinski:1999ry, Susskind:1998vk, Penedones:2010ue}.
    \item Finally, a tantalizing possibility is that the M-theory amplitude is an extremal amplitude in the S-matrix bootstrap \cite{Guerrieri:2022sod}. If so, general principles of unitarity, crossing, etc., would be enough to determine the amplitude.
\end{enumerate}
It would be nice to compare these different approaches, in particular to compute some non-protected quantity from two or more different approaches.
\\
It is sometimes said that we do not have an independent, non-perturbative bulk definition of M-theory. Nevertheless, a variant of the BFSS conjecture that seems non-perturbatively well-defined is: {\it the amplitudes of the matrix model in the BFSS limit agree with the flat space limit of the amplitudes obtained from the ABJM or the 6D (2,0) CFT.} This conjecture is still shocking as the different quantum systems {\it a priori} have very little in common.

\end{boxcomment}

\section{Other techniques for analyzing the BFSS model}
Here we give a superficial survey of various techniques that have been used to analyze the BFSS model analytically. 

\subsection{High temperature/weak 't Hooft coupling expansion}
In the high temperature limit, the effective 't Hooft coupling $\lambda \beta^3 \ll 1$ and we can develop an expansion in $\lambda_\text{eff}$. The $\beta \to 0$ limit is not quite a free theory because we are left with a non-trivial matrix integral over the Matsubara zero modes. Note that if we impose periodic boundary conditions for the fermions we obtain the IKKT matrix integral, whereas if we impose anti-periodic boundary conditions, there are no fermion zero modes. The latter case is relevant for the usual thermodynamics, see \cite{Kawahara:2007ib}.

On a similar conceptual footing, one can consider the BMN matrix model at very large mass parameter $\mu$. Then the effective 't Hooft coupling $\lambda/\mu^3 \ll 1$ and one can do ordinary Hamiltonian perturbation theory \cite{Dasgupta:2002hx} to compute the energy levels, etc.

\subsection{SUSY techniques}
The computation of the index in BFSS is perhaps the oldest case where supersymmetric techniques were applicable. We have already mentioned this in section \ref{spectrumN}; it is a subtle computation due to the flat directions. In the BMN matrix model, the flat directions of the BFSS matrix model are lifted due to the presence of mass terms. The index of the BMN model was recently computed \cite{Chang:2022mjp}.

One can also compute other quantities in the BMN matrix model using supersymmetric localization \cite{Asano:2014vba}. This technique can be used to compute correlation functions like $\ev{\Tr Z^k}$ where $Z = X_a + \i X_m$ where $a$ is an SO(6) index and $m$ is an SO(3) index. (One might ask whether these results give something non-trivial in the massless limit, but such correlation functions vanish in the BFSS vacuum due to SO(9) symmetry.) Recently, a mass-deformed version of IKKT has also been studied \cite{Hartnoll:2024csr, Komatsu:2024bop, Komatsu:2024ydh}; localization can also be used to compute the partition function and some correlation functions in this model. Both the mass-deformed IKKT and the mass-deformed BFSS model (the BMN model) have a rich set of vacua which unfortunately is beyond the scope of this review.

Beyond correlation functions, one may ask about scattering. As we have mentioned, the effective action on the moduli space is heavily constrained by non-renormalization theorems that rely on supersymmetry \cite{Paban:1998ea, Paban:1998qy, Lowe:1998vu, Sethi:1999qv}. This constrains the 4-pt amplitude (and higher-pt amplitudes), especially in the limit of zero momentum transfer in the longitudinal direction. Another recent development is the computation of the 3-pt amplitude. The 3-pt amplitude was recently computed in \cite{Herderschee:2023pza}. The 3-pt amplitude in M-theory is fixed (up to a constant) by kinematics\footnote{The 3-pt kinematics involving massless particles cannot be satisfied with real momentum. However, it can be satisfied by going to (2,9) signature, or equivalently considering the usual (1,10) signature and taking one of the spatial momenta to be imaginary.}. However, the symmetries of 11-d M-theory are not manifest in the BFSS model. Hence it is a non-trivial computation (and a test of the BFSS conjecture) that the 3-pt amplitude is reproduced by the matrix model. The idea in \cite{Herderschee:2023pza} is to uplift to a problem in 1+1D SYM compactified on a circle of circumference  $2 \pi\tilde{R}_9$. When $\tilde{R}_9$ is small, we expect this problem to reduce to the BFSS problem. The boundary conditions on the cylinder are chosen to represent the initial and final states in the 3-pt amplitude.

Now converting the scattering problem in the quantum mechanics to a field theory problem seems like making the problem harder, but actually the field theory computation has an advantage. The trick is to note that the cylinder can be interpreted as a trace in the ``open string'' channel, we are computing a supersymmetric index $\Tr ( (-1)^F e^{-2\pi \tilde{R}_9 H} \cdots )$: 
\[
\begin{tikzpicture}[>=Stealth,cylinder length/.store in=\L,cylinder length=4,cylinder x radius/.store in=\rx,cylinder x radius=0.3,cylinder y radius/.store in=\ry,cylinder y radius=1.0]
\draw[blue,thick](0,0)ellipse[x radius=\rx cm,y radius=\ry cm];
\draw[blue,thick](\L cm,0)ellipse[x radius=\rx cm,y radius=\ry cm];
\draw[blue,thick](0,\ry cm)--(\L cm,\ry cm);
\draw[blue,thick](0,-\ry cm)--(\L cm,-\ry cm);
\draw[red,line width=1.5pt](0.26 cm,0)--(\L cm-.26 cm,0);
\draw[red,->](1cm,0)..controls(1cm,0.5cm)and(0.85cm,0.7cm)..(0.75cm,\ry cm-0.1cm);
\node[below]at(0,-\ry cm){$\tau_l$};
\node[below]at(\L cm,-\ry cm){$\tau_r$};
\draw[->](0.6cm,-\ry cm-0.5cm)--node[below]{$\tau$}(1.6cm,-\ry cm-0.5cm);
\draw[black,thick,->] (\L cm+\rx cm+0.2cm,-.9 cm)
  arc[start angle=-60,end angle=60,
      x radius=\rx cm,
      y radius=\ry cm];
\node at(\L cm+1cm,0){$\widetilde R_9$};
\node at(\L cm/2,\ry cm+0.5cm){Open string channel};
\node at(\L cm+2.5cm,0){\Large $=$};
\begin{scope}[xshift=\L cm+4.5cm]
\draw[blue,thick](0,0)ellipse[x radius=\rx cm,y radius=\ry cm];
\draw[blue,thick](\L cm,0)ellipse[x radius=\rx cm,y radius=\ry cm];
\draw[blue,thick](0,\ry cm)--(\L cm,\ry cm);
\draw[blue,thick](0,-\ry cm)--(\L cm,-\ry cm);
\draw[red,line width=1.5pt,fill=red!20](\L cm/2,0)ellipse[x radius=\rx cm,y radius=\dimexpr\ry cm];
\draw[red,->](2cm,0)--(3cm,0);
\node[left]at(-0.5,0){$\langle B_L|$};
\node[right]at(\L cm+0.5cm,0){$|B_R\rangle$};
\node at(\L cm/2,\ry cm+0.5cm){Closed string channel};
\end{scope}
\end{tikzpicture}
\]
Then one can compute the index semi-classically. On the other hand, in the closed string channel, one views the same quantity as an overlap between the supersymmetric boundary states $\bra{B_L}$ and $\ket{B_R}$, which are chosen to be D1 cousins of the block diagonal states described around \eqref{blockScatter}. The result of the computation agrees with the BFSS prediction.

\section{Matrix Bootstrap}

\subsection{Bootstrapping equations of motion}
We will now turn to a particular non-perturbative approach to solving large $N$ matrix systems, including the BFSS matrix model. This approach is sometimes called the ``matrix bootstrap''; its main features are 
\begin{enumerate}
    \item It is non-perturbative in the coupling; it does not rely on any weak-coupling expansion.
    \item It works directly in the 't Hooft large $N$ limit. 
    \item It can be used to solve quantum systems, including those that have a sign problem in Euclidean signature.
    \item It works even for systems that are metastable; e.g., systems which are strictly speaking only well-defined at infinite $N$. %
\end{enumerate}
Some potential downsides of the method are that it only produces inequalities on interesting observables. In some simple contexts it has been argued that these inequalities will be enough to give islands that converge to the exact solution \cite{Lin:2020mme}; but for most situations there is no guarantee of convergence. Of course, according to the bootstrap philosophy, rigorous inequalities might still be interesting, especially if there are ``kinks'' in the allowed region. Another potential downside is that for a generic system with $D$ matrices, the number of single trace operators of length $L$ grows  $\sim D^L$, which can quickly become computationally expensive.

The simplest context is the single matrix integral \cite{Anderson:2016rcw,Lin:2020mme}. The goal is to compute moments 
\begin{align}
    \langle \tr X^k \rangle = \lim_{N \to \infty} \frac{1}{Z_N} \int \d X \, \mathrm{tr} ( X^k)  e^{-N^2 \tr V(X)} .
\end{align}
Here $V(X)$ is a polynomial in $X$. We have normalized $\tr 1 = 1$, e.g., $\tr = \frac{1}{N} \Tr$.
At infinite $N$ (in the 't Hooft limit), we can use large $N$ factorization to reduce multi-trace observables to single traces.
To bootstrap, we need a set of consistency conditions and a set of positivity constraints. For consistency conditions, we will use the Schwinger-Dyson or loop equations:
\begin{align} \label{eq:loop}
&\begin{tikzpicture}[baseline={(current bounding box.center)},
    scale=0.9,
    blob/.style={
        circle, 
        draw, 
        pattern=north east lines, 
        minimum size=1.5cm, 
        inner sep=0pt
    },
    smallblob/.style={
        circle,
        draw,
        pattern=north east lines,
        minimum size=1.0cm,
        inner sep=0pt
    },
    vertex/.style={
        circle,
        fill=black,
        inner sep=1.5pt
    },
    line/.style={draw,
        double,
        double distance=1.5pt},
    arrowline/.style={
        draw, double, double distance=1.5pt,
        decoration={markings, mark=at position 0.65 with {\arrow[scale=0.7]{Stealth}}},
        postaction={decorate}
    },
]
\node[blob] (lhs_blob) at (0,0) {};
\draw[arrowline] (-2, 0) -- (lhs_blob.180);
\foreach \angle in {120, 90, 60, 0, -60, -90, -120} {
    \draw[line] (lhs_blob.\angle) -- ++(\angle:1.1cm);
}
\node at (3, 0) {$=$};
\node at (7.5, 0) {$+$};
\draw[arrowline] (4.2, 0) -- (6.5, 0);
\node[smallblob] (top_blob) at (5.5, 1.5) {};
\draw[line] (top_blob.180) -- ++(-0.7,0);
\draw[line] (top_blob.45) -- ++(45:0.7cm);
\draw[line] (top_blob.-45) -- ++(-45:0.7cm);
\node[smallblob] (bottom_blob) at (5.5, -1.5) {};
\draw[line] (bottom_blob.180) -- ++(-0.7,0);
\draw[line] (bottom_blob.45) -- ++(45:0.7cm);
\draw[line] (bottom_blob.-45) -- ++(-45:0.7cm);
\node[blob] (rhs_blob) at (11, 0) {};
\foreach \angle in {120, 90, 60, 0, -60, -90, -120} {
    \draw[line] (rhs_blob.\angle) -- ++(\angle:1.1cm);
}
\node[vertex] (v) at (9.4, 0) {};
\draw[arrowline] (8.2, 0) -- (v);
\draw[line] (v) -- (rhs_blob.170); %
\draw[line] (v) -- (rhs_blob.190); %
\end{tikzpicture}\\
    & \hspace{1.8cm}  \langle{\tr X^k V'(X)}\rangle \; = \;  \sum_{l=0}^{k-1} \langle{\tr X^\ell}\rangle \langle{\tr X^{k-\ell-1}}\rangle %
\end{align}
In writing the double trace as a product of two single traces, we have assumed large $N$ factorization, and {\it hence we have input infinite $N$ into the bootstrap}. For a polynomial potential $V$, we may view this equation as a recursion relation between lower moments and higher moments. In \eqref{eq:loop} we have depicted the Schwinger-Dyson equations for $V = \hf X^2 + \frac{1}{3} g X^3$.

Then we combine this with positivity of the Hankel matrix 
\begin{align}
    \mathcal{M}_{i,j} = \langle {\tr X^{i+j} } \rangle, \quad 
    \mathcal{M} \succeq 0.
\end{align}
The notation $\mathcal{M} \succeq 0$ indicates that $\mathcal{M}$ is a positive semi-definite matrix, e.g., that all of its eigenvalues are non-negative. An equivalent characterization is that $p^\dagger \mathcal{M} p \ge 0$ for all vectors $p$. This positivity follows from the fact that we may consider a general polynomial in the matrices
\begin{align}
    p(X) = \sum_k p_k X^k, \quad \ev{\bar{p}(X)p(X) } \ge 0 \Leftrightarrow \sum_{ik} \bar{p}_i \mathcal{M}_{ik} p_k \ge 0,
\end{align}
where we have used the fact that $X$ is a Hermitian matrix.

\begin{figure}
    \centering
    \includegraphics[width=0.7\linewidth]{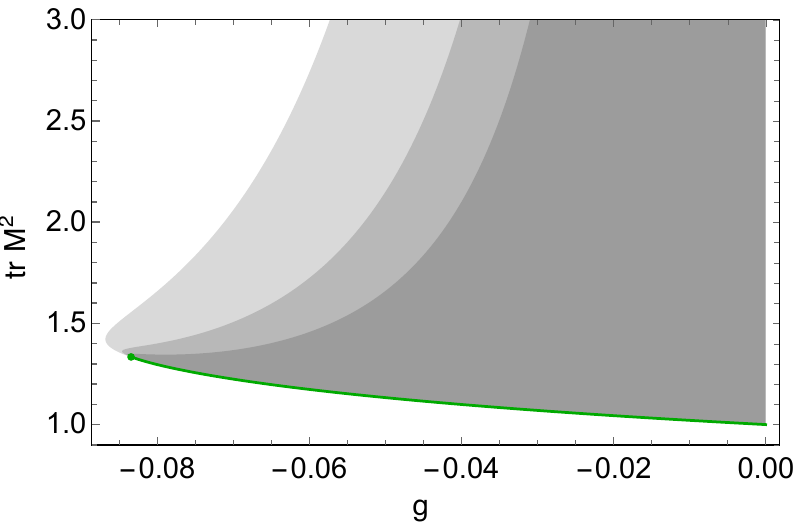}
    \caption{1-matrix model bootstrap for unbounded potential $V = \frac12 \tr M^2 + \frac{g}{4} \tr M^4 $. The {\color{dgreen} green curve} is the exact large $N$ solution. The shaded regions are the allowed regions from the bootstrap; as we increase the constraints the tip of the peninsula rapidly approaches the critical value of $g \to g_\text{c} = -1/12 \approx -0.083$. This is also the radius of convergence for the 't Hooft expansion. }
    \label{fig:1matrix}
\end{figure}

\begin{exercise}{1-matrix model bootstrap}
Consider $V = \hf X^2 + \frac{1}{3} g X^3$. Construct the Hankel matrix $\mathcal{M}_{ij}$ up to length $i+j \le L$. Use the loop equations \eqref{eq:loop} to express all elements as a function of $\tr X$ and $g$. Then use \texttt{SemidefiniteOptimization[]} in Mathematica to generate a plot of the allowed region for the $\tr X$ as a function of $g$. At large values of $L$, one should be able to estimate the range of $g$ where the problem is well-defined in the $N \to \infty$ limit.
\end{exercise}

\begin{exercise}{Single variable integral bootstrap}
For an even easier exercise, derive the Schwinger-Dyson equation for the measure over a single bosonic variable:
\begin{align}
 \langle x^n \rangle = {Z}^{-1}  \int \d x  \, x^n e^{-V(x)}
\end{align}
where $V$ can be your favorite bounded polynomial, e.g., $V(x) = \hf x^2 + \qrt g x^4$. Define again the Hankel matrix $\mathcal{M}_{i,j} = \langle {x^{i+j} } \rangle$ and use \texttt{SemidefiniteOptimization[]} to bound the correlation functions. Can you get a high precision estimate of $\langle x^2 \rangle$, say at $g=1$?
\end{exercise}

In principle, large $N$ factorization means that different correlators are related by a set of quadratic equations\footnote{For finite $N$, one can instead impose trace relations \cite{Kazakov:2024ool}.}. This means that the semi-definite program is non-linear. For some simple problems like the 1-matrix model, one can simply scan over a small number of variables. However, more generally, one needs to use the method of non-linear relaxation  \cite{Kazakov:2022xuh}. The idea is to replace $q = x^2$ with $q \ge x^2$. This can be encoded in a positive semi-definite matrix:
\begin{align}
    \mathcal{M}_\text{NLR} = \left(\begin{array}{cc}
1  & x \\
x & q \\
\end{array}\right) \succeq 0.
\end{align}
More generally, let $\vec{x}$ be a set of single trace variables, e.g., $x_k = \ev{\tr X^k}$. Then we define the matrix of variables $Q_{ij}$. To perform non-linear relaxation, we replace all occurrences of $x_i x_j$ in the loop equations with $x_i x_j \to Q_{ij}$. Then we enforce
\begin{align}
    \mathcal{M}_\text{NLR} = \left(\begin{array}{cc}
1  & \vec{x} \\
\vec{x}^T & Q \\
\end{array}\right) \succeq 0.
\end{align}
One can also impose a slightly more refined constraint \cite{Lin:2025srf} 
$\mathcal{M} \succeq Q,$ which follows from positivity of the covariance matrix $\langle\operatorname{tr}\left(\mathcal{O}_i-\left\langle\mathcal{O}_i\right\rangle\right)^{\dagger}\left(\mathcal{O}_j-\left\langle\mathcal{O}_j\right\rangle\right)\rangle=\mathcal{M}_{i j}-Q_{i j} \succeq 0 .$

One can generalize the bootstrap approach to consider multi-matrix integrals, see \cite{Lin:2020mme} and \cite{Kazakov:2021lel}. One can consider all observables that are single trace words made up of $L$ letters. The number of such observables grows rapidly, $\sim D^L$. However, the number of independent loop equations also grows exponentially. Furthermore, one should generalize $\mathcal{M}$ to be the inner product matrix of arbitrary words. This leads to a positivity matrix that is also exponentially large. Despite these challenges, \cite{Kazakov:2021lel} achieved high precision estimates for simple correlation functions in a certain 2-matrix model with an interaction of the form $V = V(X) + V(Y) - g^2 \,  \tr  [X,Y]^2$ that is not believed to be solvable by analytic means. To achieve these impressive high precision estimates, it is important to use the non-linear relaxation method. Unlike the 1-matrix model, as one considers larger and larger values of $L$, the number of unknown variables grows and it is impractical to scan over all such variables.

\subsection{Quantum mechanical bootstrap}
Now we turn to the generalization from the matrix integral to matrix quantum mechanics. The key insight \cite{Han:2020bkb} is to work in the Hilbert space/Hamiltonian formalism. One replaces the Schwinger-Dyson equations with the equations of motion:
\begin{align}
    \ev{[H,\mathcal{O}]} = 0
\end{align}
where $\mathcal{O}$ is any operator and the expectation value is with respect to an energy eigenstate (or, more generally, a density matrix which commutes with the Hamiltonian).

The positivity constraints are again of the form $\mathcal{M}_{ij} = \langle  O^\dagger_i O_j \rangle, \; \mathcal{M} \succeq 0 $. This positivity constraint only relies on the positivity of the Hilbert space inner product; it holds for any quantum-mechanical system. For Euclidean problems, we are using reflection positivity and {\it not} positivity of the Euclidean measure; fermionic degrees of freedom are not a problem for this quantum mechanical bootstrap. 

 One can also scan over $\ev{H} = E$. For a simple quantum system like a non-relativistic particle in a potential $V(x)$, one can scan over $E$ while minimizing/maximizing over some simple 1-pt function like $\ev{x^2}$. If one plots $E$ vs $\ev{x^2}$, typically one finds an ``archipelago'' where there are a few islands corresponding to the lowest eigenvalues of the Hamiltonian; at larger values of energy there is a peninsula where individual eigenstates cannot be resolved.

\begin{exercise}{Bootstrapping simple quantum systems}
One can apply the quantum mechanical bootstrap to simple quantum systems, like a particle in a potential (see e.g. \cite{Berenstein:2021dyf, Berenstein:2022unr}). Compute the ground state energy of the quantum anharmonic oscillator using these ideas.

A more interesting problem is to consider the ``toy supermembrane.'' This is a non-relativistic particle in 2 dimensions with potential $V = x^2 y^2$. Classically this model has some flat directions and was considered a toy model for D0 brane quantum mechanics \cite{Douglas_1997, Komatsu:2024vnb}. 
\end{exercise}

For the application to large $N$ matrix quantum mechanics, we should consider the equations of motion obtained from single trace operators $    \ev{[H,\Tr \mathcal{O}]} = 0$ and the corresponding positivity matrix of adjoints $\ev{\Tr  O^\dagger_i O_j} \ge 0 $. Note that the commutator of two single traces gives a single trace. 

In addition we use cyclicity of the trace $+$ canonical commutation relations\footnote{Here $\tr$ is over the SU($N$) indices and should {\it not} be confused with the trace over the quantum Hilbert space. This is why we cannot simply write $\tr (AB) = \tr(BA)$.}, e.g.,
    \begin{align} \label{eq:exampleCyc}
       \ev{ \tr X^{I_1}  X^{I_2} X^{I_3} P^{I_4} X^{I_5} X^{I_6}}  = \ev{\tr   X^{I_2} X^{I_3} P^{I_4} X^{I_5} X^{I_6} X^{I_1}} + \i \ev{\tr X^{I_2} X^{I_3}} \ev{\tr X^{I_5} X^{I_6}} \delta^{I_1 I_4} 
    \end{align}
In the Hamiltonian approach, these cyclicity relations are the only ingredient that generates double trace relations, and therefore the only place where we input $N= \infty$.

\subsection{Finite temperature}
The above constraints allow us to bootstrap systems in an energy eigenstate, or in the large $N$ limit, the microcanonical ensemble. However, we would also like to bootstrap systems in the canonical ensemble. To do so, one leverages an inequality that is sometimes called the energy-entropy balance (EEB) inequality (Araki and Sewell \cite{Araki:1977px}, see also \cite{Fawzi:2023fpg}). The EEB inequality states that for any operator $O$,
\begin{align} \label{eeb}
 \log \frac{\ev{O^\dagger O}}{ {\ev{O O^\dagger } } }\le  \beta \frac{ \ev{O^\dagger [H,O] } }{ \ev{O^\dagger O}} 
\end{align}
where expectation values are taken with respect to the thermal state $\rho = \frac{1}{Z} e^{-\beta H}$.
Together with time-translation invariance $\ev{[O,H]} =0$, positivity $\ev{O^\dagger O} \ge 0$, and normalization of the trace $\ev{1} = 1$, the EEB inequality for all operators $O$ is equivalent to the KMS condition. %
Recently this has been used to bootstrap the thermodynamics of the ungauged 1-matrix model and the 2-matrix model with Yang-Mills like interaction, see \cite{Cho:2024kxn}. For this application, one chooses $O$ to be an operator that transforms in the adjoint of SU($N$) or U($N$). A derivation of this inequality using log-convexity of the thermal 2-pt function may be found in \cite{Cho:2025euc}, see also Appendix A of \cite{Cho:2024owx} and Theorem 5.3.15 of \cite{Bratteli:1996xq}. 

A simple case of the EEB inequality is to take $\beta\to \infty$; then we derive the {\it ground state positivity} condition:
\begin{align}
    \frac{ \ev{O^\dagger [H,O] } }{ \ev{O^\dagger O}} = \bra{O} H \ket{O} - E_0 \ge 0 , \quad \ket{O} = \frac{ O \ket{\Omega}}{\bra{\Omega} O^\dagger O \ket{\Omega}^{1/2}}.
\end{align}
This inequality says that perturbing the ground state by acting with an operator $O$ may only increase the energy of the system. A slight generalization of this statement is that the matrix
\begin{align} \label{gsp}
    \mathcal{N}_{ij} =\ev{O^\dagger_i [H,O_j] }, \quad \mathcal{N} \succeq 0.
\end{align}
This follows from the requirement that a generic superposition of perturbations $c_i O_i$ acting on the ground state $\ket{\Omega}$ raises the energy. This inequality can be used to obtain very precise bounds on properties of the ground state \cite{LinZheng2}. It is a bit simpler to work with in practice, since we do not have to deal with the non-linear nature of \eqref{eeb}, although this can be dealt with using convex relaxation \cite{Fawzi:2023fpg, Cho:2024kxn}.
\begin{exercise}{Energy-entropy inequality }
    Consider the harmonic oscillator at finite temperature\footnote{I thank everyone at an SITP lunch for suggesting/discussing this example.}. Consider the  inequality \eqref{eeb} with $O = a^n$. Then separately consider $O = (a^\dagger)^n$. Argue that for both sets of inequalities to be true, we must have saturation
    \begin{align}
        \log \frac{\ev{O^\dagger O}}{ {\ev{O O^\dagger } } } =   \beta \frac{ \ev{O^\dagger [H,O] } }{ \ev{O^\dagger O}} 
    \end{align}
    Use this to argue that the density matrix $\rho \propto \exp(- \beta H)$.
\end{exercise}

An interesting recent development is that the bootstrap can be generalized to time-dependent problems:
\begin{boxcomment}{Time dependent bootstrap}
 Here we summarize the LMN bootstrap method for time-dependent 1-pt functions \cite{Lawrence:2024mnj}.
The primal problem is:
\begin{align}
\operatorname{minimize} & \operatorname{Tr} \mathcal{O} \mathcal{M}(T) \\
\text { subject to } & \mathcal{M}(t) \succeq 0 \label{posM} \\
& \operatorname{Tr} A^{(i)} \mathcal{M}(t)= a^{(i)} \label{constrA}\\
& \operatorname{Tr}\left(D^{(k)}-C^{(k)} \frac{\d}{\d t}\right) \mathcal{M}(t)=0 \label{constrB}\\
& \mathcal{M}(T=0) = \mathcal{M}_0\label{constrInit}
\end{align}
Here $\mathcal{M}(t)$ is a matrix that encodes a set of Lorentzian 1-pt functions. The constraint \eqref{constrB} encodes the Heisenberg equations of motion and \eqref{constrInit} is the initial condition (where information about the initial state is input). \eqref{posM} inputs Hilbert space positivity, and \eqref{constrA} enforces relations between 1-pt functions due to the canonical commutation relations. 

The dual problem is
\begin{align}
\text { maximize } & \lambda_D^{(k)}(0) \operatorname{Tr} C^{(k)} \mathcal{M}_0  \label{dualOpt} \\
\text { subject to } & \lambda_D^{(k)}(T) C^{(k)}= \mathcal{O} \label{dualC1}\\
& \lambda_A^{(i)}(t) A^{(i)}+\left(D^{(k)}+C^{(k)} \frac{\d}{\d t}\right) \lambda_D^{(k)} \succeq 0 .\label{dualC2}
\end{align}
To derive the dual problem, one introduces an action for the primal problem, with Lagrange multipliers $\lambda_A(t), \lambda_D(t)$ which enforce \eqref{constrA} and \eqref{constrB}. One also introduces a Lagrange multiplier $\Lambda(t) \succeq 0$ that enforces \eqref{posM}. Then one can integrate out (e.g. enforce the equations of motion) for the $\mathcal{M}(t)$ variable, which leads to \eqref{dualC1} and \eqref{dualC2}. The resulting action \eqref{dualOpt} will then just involve the Lagrange multiplier evaluated at $t=0$.

The main point is that in the dual problem, {\it we do not need to solve the Heisenberg equations of motion}, which is generically impossible in a strongly coupled system. We have managed to convert the equations into a set of inequalities on the dual variables $\lambda(t)$. Roughly speaking, one can expand $\lambda(t) = \sum_i \lambda_i e_i(t)$ over some finite basis of functions $e_i(t)$ and try to optimize over the choice of coefficients $\lambda_i$ subject to the constraints. As long as we find a solution to the dual problem (a feasible function that satisfies \eqref{dualC2}), we immediately derive a bound on the primal problem. In particular, we do not need to search over an infinite-dimensional basis of functions $e_i(t)$ to derive a bound! 

One can generalize this method to solve a variety of time-dependent problems, including the 2-pt function in Euclidean signature \cite{Cho:2025euc}. This allows one to also study thermal properties by imposing the KMS condition on 2-pt functions without using the EEB inequalities \cite{Cho:2025euc}.
\end{boxcomment}

\subsection{Application to BFSS}
We now discuss the application of these bootstrap ideas to BFSS. Optimistically, using the above ideas, one should be able to compute a wide variety of quantities that probe the rich physics of this model that we have reviewed. For example, one could hope to compute the thermodynamics of the model, which would test the Bekenstein-Hawking area formula and go beyond into the stringy black hole regime, as we discussed in \eqref{therm}. For now, however, we will focus on a basic but surprisingly non-trivial question:
what is the size of the bound state wavefunction? This is non-trivial since this is a low-energy (and therefore strong coupling) property of the BFSS system that cannot be computed using any known weak-coupling method.
In the original BFSS paper \cite{Banks:1996vh}, two estimates were given. One estimate was $r/\ell_p \sim N^{1/9}$, which is smaller than the size of the M-theory bubble $r/\ell_p \lesssim N^{1/7}$. Another estimate, later confirmed by Polchinski \cite{Polchinski:1999br}, is the much larger value $r/\ell_p \gtrsim N^{1/3}$, which is the answer expected from 't Hooft scaling\footnote{To see this, note that the engineering dimension $[X] = \text{mass}$, so $1/ N  \ev{\Tr X^2}  \lambda ^ {-2/3} \sim r^2/N^{-2/3} \sim  O(1)$. }. From a modern perspective \cite{Lin:2023owt}, we can recognize \cite{Polchinski:1999br}'s arguments as a bootstrap proof. (For an intuitive account of Polchinski's argument, see \cite{Susskind:1998vk}.) It is based on  exactly the same constraints and positivity ideas that were listed in \cite{Han:2020bkb} and outlined above. These are: 
\begin{boxc}{Constraints:}
    \begin{align} 
    &\ev{ [H,\Tr X^2]} = 0 \RA \ev{\Tr X^I P_I} + \ev{\Tr P^I X_I } =0. \label{xx}\\
    &\ev{\Tr [X^I,P^I]} = 9 \i N^2.  \label{xp}\\
    &\ev{[H, \Tr X P]} = 0 \RA  -2\ev{K} + 4 \ev{V} + \ev{F} = 0. \label{hxp}\\
    & \ev{H} = E \Rightarrow \ev{K} + \ev{V} + \ev{F} = E . \label{h}
    \end{align}
\end{boxc}
Combining \eqref{xx} and \eqref{xp} gives $\ev{\Tr X^I P^I} =  9 \i N^2/2$. Combining \eqref{hxp} and \eqref{h}, we may eliminate $\ev{F}$:
\begin{align}    2\ev{K} = \tfrac{2}{3}E + 2\ev{V}. \label{virGround}
\end{align}
Now we list some relevant positivity constraints:
\begin{boxc}{Positivity}
\begin{align}
&\left(
\begin{array}{cc}
\ev{\Tr X^2}  & \ev{\Tr X P} \\
\ev{\Tr  P X} & \ev{\Tr P^2} \\
\end{array}
\right) \succeq 0, \quad 
\RA  \sum_I \ev{ \operatorname{Tr} X^2 } \ev{ \operatorname{Tr}\left(P^{I} P^{I}\right) } \geq \frac94 N^{4}. \label{xxpp}\\
&\left(
\begin{array}{cc}
\Tr X^4  & \Tr X^2 Y^2 \\
\Tr   X^2 Y^2 & \Tr X^4 \\
\end{array}
\right) \succeq 0, \quad 
\left(
\begin{array}{cc}
\Tr X^2 Y^2  & \Tr XYXY \\
\Tr   XYXY & \Tr X^2 Y^2 \\
\label{xxyy}
\end{array}
\right) \succeq 0,\\
& \left(\begin{array}{cc}
\langle\operatorname{Tr} 1\rangle & \left\langle\operatorname{Tr} X^2\right\rangle \\
\left\langle\operatorname{Tr} X^2\right\rangle & \left\langle\operatorname{Tr} X^4\right\rangle
\end{array}\right) \succeq 0, \label{x4pos}
\end{align}
\end{boxc}
We start with the RHS of \eqref{xxpp}, which is essentially the uncertainty principle for matrices. 
We use \eqref{virGround} to replace $\ev{\Tr P^I P^I}$ (kinetic energy) with potential energy:%
\begin{align}
    2\ev{V} = - \frac{1}{2g^2_\ym} \ev{\Tr [X^I, X^J]^2} = - \frac{9 \times 8 }{2 g^2_\ym } \ev{ \Tr [X,Y]^2} = \frac{72 }{ g^2_\ym } \ev{ \Tr X^2 Y^2 - \Tr XYXY }.
\end{align} 
We have used SO(9) symmetry to focus on just 2 of the matrices, which we call $X$ and $Y$. We can then relate both terms to $\ev{\Tr X^4}$ using the positivity relations \eqref{xxyy}.
Finally, we can replace $\ev{\Tr X^2}$ with $\ev{\Tr X^4}$ using \eqref{x4pos}.
This allows us to write the inequality in terms of just $\ev{\Tr X^4}$:
\begin{align} \boxed{\ev{N \Tr X^4}^{1/2}   \left( \frac{144}{g^2_\ym}\ev{\Tr X^4}+\frac{2 E}{3} \right)   \geq   \frac94 g^2_\ym N^4 } \end{align}
Let us make a few comments. Setting $E=0$ recovers Polchinski's point. Assuming parametric saturation of the bound implies that ``typical eigenvalue'' $r \sim \lambda^{1/3}$, which is roughly the size of the Type IIA supergravity region. The scale at which the bound varies is $E/N^2 \sim \lambda^{1/3}$. Recall that the regime of validity of supergravity is $E/N^2 \ll \lambda^{1/3}$, see e.g. \eqref{et-thermo}. It is interesting that these simple arguments give us non-trivial dynamical information even in the strong coupling regime. %

Recognizing the Polchinski virial theorem as a bootstrap bound immediately allows us to improve it. In particular, we can derive a bound on $\ev{\tr X^2}$ which cannot be obtained from the above arguments alone. To do so, we should include information about the fermionic term $F$. The idea is to consider a $3 \times 3$ positivity matrix corresponding to the operators $O_I, P_I, X_I$ where we have written the fermionic term as $\Tr O_I X_I$.
\begin{align}
    \mathcal{M}_3=\left[\begin{array}{ccc}
\frac{1}{9}\left\langle\operatorname{Tr} O_I O_I\right\rangle & \frac{2}{9}\left(\frac{1}{3} E-\langle V\rangle\right) & 0 \\
\frac{2}{9}\left(\frac{1}{3} E-\langle V\rangle\right) & \left\langle\operatorname{Tr} X^2\right\rangle & \frac{\mathrm{i} }{2} N^2 \\
0 & -\frac{\mathrm{i} }{2 } N^2 & \frac{2}{9 g^2_\ym }\left(\frac{1}{3} E+\langle V\rangle\right)
\end{array}\right] \succeq 0 .
\end{align}
In the next section, we show\footnote{The bound in \cite{Lin:2023owt} was improved in \cite{LinZheng1} by treating the fermionic constraints more systematically.} that $1/9 \ev{\Tr O_I O_I }\le 64 N^2/3$, see \eqref{oioi}. Then demanding that $\det \mathcal{M}_3 \ge 0$ gives %
\begin{align} \label{bdXX}
(\tfrac13 E +  V) \left(48 N^2 \ev{\Tr X^2}-(\tfrac13 E -  V)^2  \right) \ge 54 g^2_\ym  N^6,\\
\frac{1}{N \lambda^{2/3}} \ev{\Tr X^2} \ge \frac{3}{16} 3^{1/3} \approx 0.27 , \quad E=0
\end{align}

\begin{boxc}{Improving Polchinski's bound}
   
   With a little more work \cite{LinZheng1}, one can derive the slightly better level 6 analytic bound
   \begin{align}
       \frac{1}{N (g^2_\ym N)^{2/3}}  \left\langle\operatorname{Tr} {X}^2\right\rangle = \langle\tr \tilde{X}^2\rangle \geq \frac{3}{4}\left(\frac{3}{50}\right)^{1 / 3} \approx 0.2936
   \end{align}

   This should be compared to the Monte Carlo value of 0.378 \cite{Pateloudis:2022ijr}. It is remarkable that these simple analytical bounds can recover at least $80\%$ of the Monte Carlo result.

   This result implies that $r/\ell_p \gtrsim N^{1/3}$, which is the size of the supergravity region, see exercise \eqref{thermodynamicsEx}.
\end{boxc}

\begin{boxcomment}{Intuition and anti-intuition about the bound state size}
There is an intuitive, non-rigorous argument \cite{Susskind:1998vk} that the wavefunction has a size $r/\ell_p \sim N^{1/3}$ that is closely related to the above ideas.
Consider some matrix $X_1$, it should have a typical size $\tr X_1^2 \sim N r^2$, where $r$ is a typical eigenvalue of $X_1$. Then consider a different matrix $X_2$. The idea is to view the off-diagonal elements of the matrix $X_2$ as harmonic oscillators with frequency given by our previous estimate \eqref{eq:oscfreq} $\omega \sim R r/\ell_p^3$. If we assume that each matrix element is in the oscillator ground state, the typical size of an off-diagonal matrix element $(\Delta X_{ij})^2 \sim \ell_p^3 / r$. Hence $\langle{\Tr (X_2)^2\rangle} \sim N^2 \ell_p^3/r$. But rotational invariance implies $\langle{\Tr (X_1)^2\rangle} = \langle{\Tr (X_2)^2\rangle} \RA  r^3 \sim N \ell_p^3 $. This gives the desired estimate. Of course, the ground state wavefunction of the BFSS theory is the wavefunction of a strongly coupled many-body system; to replace $\sim$ with precise inequalities, we must use the bootstrap arguments. \\

While the above argument provides some intuition for why the wavefunction is big, the large size of the bound state is counterintuitive from the point of view of the BFSS scattering conjecture.  Here we quote\footnote{The quote has been lightly edited to correct typos and to adopt our notation.} from Susskind \cite{Susskind:1998vk}, who puts it colorfully:
\begin{quote}
[T]he history of the scattering process has two very different but equivalent descriptions. In the usual spacetime supergravity description two small particles come in from infinity and remain essentially non-interacting until they come within a distance of order $\ell_p$. They interact for a short time and then separate into final particles which cease to interact as soon as they
are separated by $\ell_p$. In light cone units the interaction lasts for a time $\ell_p N / (|\vec{p}| R)$. The holographic matrix description also begins with asymptotically distant non-interacting objects. In this description the constituents begin to
merge and interact when their separation is of order $N^{1/3} \ell_p$. As they approach, the many body wave function begins to more and more resemble the ground state. The system remains in this entangled state for a light cone time of order $\ell_p N^{4/3}/(|\vec{p}| R)$  and then separates into non-interacting final clusters. The situation is particularly perplexing if the energy is not very large and the impact parameter is much larger than $\ell_p$. [Then in the] gravity description the particles miss each other and just continue without significant deflection. {\it Exactly how this miracle happens from the SYM description is still a mystery.}
\end{quote} 
A related question is: to what extent can one isolate the low energy degrees of freedom in the matrix model that encodes the $M$-theory region?
\end{boxcomment}

Before moving on, let us note that for any supersymmetric quantum mechanical system $\{ Q_\alpha, Q_\beta\} \propto \delta_{\alpha \beta} H$ one can directly bootstrap a SUSY invariant state $Q_\alpha \ket{\Omega}= 0$. This is done by writing $\ev{Q_\alpha O} = \ev{O Q_\alpha} = 0$. Since a SUSY invariant state is automatically a ground state of $H$, we may further assume that it preserves any global symmetries, including the $R$-symmetry under which $Q$ transforms. At large $N$, the useful condition becomes:
\begin{boxc}{SUSY ground state bootstrap}
    For a supersymmetric quantum system with at least one ground state that preserves SUSY, one can instead use the supercharge equations of motion:  
    \begin{align}
    \bra{\Omega} \{Q_\alpha, O_\alpha\} \ket{\Omega} = 0.
    \end{align}
    Here $O_\alpha$ is any fermionic operator. (Since the ground state preserves the $R$-symmetry, we get useful equations by forming singlets under the $R$-symmetry.)
    Together with the usual positivity matrices, it can be shown \cite{LinZheng1} that ground state positivity \eqref{gsp} is automatically implied by the supercharge constraints.
\end{boxc}

\subsubsection{Decomposing into SO(9) blocks}

SO(9) symmetry implies that the only ground state correlation functions are SO(9) invariants. However, to derive positivity constraints, we must consider operators that transform non-trivially under SO(9) in intermediate steps. For example, to prove that $\ev{\tr X^I X^I} \ge 0$, we observe that it is the norm-squared of the operator $X^I$, which is an SO(9) vector. 

More generally, the positivity matrix $\mathcal{M}$ can be written as a direct sum of block-diagonal matrices, with each block associated to some irrep of SO(9). Given that the BFSS matrix model has 9 bosonic matrices and 16 fermionic matrices, we would like to avoid explicitly enumerating the operators. The situation is somewhat similar to how one performs the conformal block decomposition in the conformal bootstrap. 

To illustrate the general procedure, let us consider the 4-fermion correlator
\begin{align}
    \mathcal{M}_{\alpha \beta, \eta \epsilon} = \ev{\tr(\psi_\alpha \psi_\beta \psi_\eta \psi_\epsilon)}.
\end{align}
We can view $\mathcal{M}$ as a giant $16^2 \times 16^2$ positivity matrix indexed by $i = (\alpha, \beta), j = ( \eta, \epsilon)$. We would like to avoid explicitly constructing such a large matrix. To this end, note that we can fuse $\alpha, \beta$ into some irrep $R$. Since the ground state is SO(9) invariant, we must also fuse $\eta, \epsilon$ into the irrep $\bar{R} = R$ to get a singlet. This means:
\begin{align}
\label{4fermion}
        \ev{\tr(\psi_\alpha \psi_\beta \psi_\eta \psi_\epsilon)}
    &={\red a_1} \delta_{\alpha \beta} \delta_{\eta \epsilon}+{\red a_9} \gamma_{\alpha \beta}^I \gamma_{\eta \epsilon}^I+ {\red a_{36}} \gamma_{\alpha \beta}^{I J} \gamma_{\eta \epsilon}^{I J}+ {\red a_{84}} \gamma_{\alpha \beta}^{I J K} \gamma_{\eta\epsilon}^{I J K}+{\red a_{126}} \gamma_{\alpha \beta}^{I J K L} \gamma_{\eta\epsilon}^{I J K L}
\end{align}
We have written down the 5 possible irreps that can be obtained by fusing two spinors. Thus we have already reduced the $16^2 \times 16^2$ matrix to just 5 unknown variables.
Now consider cyclicity of the trace (together with the anti-commutation relations for the fermions) $\tr(\psi_\alpha \psi_\beta \psi_\eta \psi_\epsilon) = -\tr( \psi_\beta \psi_\eta \psi_\epsilon \psi_\alpha) + \frac{1}{2} \delta_{\alpha\beta}\delta_{\eta\epsilon}+\frac{1}{2}\delta_{\alpha\epsilon}\delta_{\eta\beta}$. This gives the alternative expression %
\begin{align} \la{cyclicity4f}
    {\red a_1} \delta_{\alpha \beta} \delta_{\eta \epsilon}+{\red a_9} \gamma_{\alpha \beta}^I \gamma_{\eta \epsilon}^I+ {\red a_{36}} \gamma_{\alpha \beta}^{I J} \gamma_{\eta \epsilon}^{I J}+ {\red a_{84}} \gamma_{\alpha \beta}^{I J K} \gamma_{\eta\epsilon}^{I J K}+{\red a_{126}} \gamma_{\alpha \beta}^{I J K L} \gamma_{\eta\epsilon}^{I J K L} = \\
    \frac{1}{2} \delta_{\alpha\beta}\delta_{\eta\epsilon}+\frac{1}{2}\delta_{\alpha\epsilon}\delta_{\eta\beta}  - {\red a_1} \delta_{ \beta \eta} \delta_{\epsilon \alpha}-{\red a_9} \gamma_{ \beta \eta}^I \gamma_{\epsilon \alpha}^I- {\red a_{36}} \gamma_{ \beta \eta}^{I J} \gamma_{\epsilon \alpha}^{I J}-{\red a_{84}} \gamma_{ \beta \eta}^{I J K} \gamma_{\epsilon \alpha}^{I J K}- {\red a_{126}} \gamma_{ \beta \eta}^{I J K L} \gamma_{\epsilon \alpha}^{I J K L} 
\end{align}
To solve this equation, we need the crossing relations for the SO(9) blocks. We have introduced a graphical notation, where each vertex corresponds to a Clebsch-Gordan symbol, and internal lines represent sums over common indices. We want to expand an $s$-channel block in a basis of $t$-channel blocks:
\begin{align}
\begin{tikzpicture}[baseline={(current bounding box.center)},scale=0.9]
\draw[thick] (-1,1) node[left] {$\beta$} -- (0,0);
\draw[thick] (-1,-1) node[left] {$\alpha$} -- (0,0);
\draw[thick] (0,0) -- (1.5,0) node[midway, above] {$R_s$};
\draw[thick] (1.5,0) -- (2.5,1) node[right] {$\eta$};
\draw[thick] (1.5,0) -- (2.5,-1) node[right] {$\epsilon$};
\end{tikzpicture}
\quad = \quad \sum_{R_t} \; \mathbb{F}_{R_s,R_t}\left[\begin{array}{ll}
{\bf 16} & {\bf 16} \\
{\bf 16} & {\bf 16}
\end{array}\right]  \;
\begin{tikzpicture}[baseline={(current bounding box.center)},scale=0.9]
\draw[thick] (-1,1) node[left] {$\beta$} -- (0,0);
\draw[thick] (0,0) -- (1,1) node[right] {$\eta$};
\draw[thick] (0,0) -- (0,-1.25) node[midway, right] {$R_t$};
\draw[thick] (-1,-2.25) node[left] {$\alpha$} -- (0,-1.25);
\draw[thick] (0,-1.25) -- (1,-2.25) node[right] {$\epsilon$};
\end{tikzpicture}
\la{defCrossing}
\end{align}
Here $\mathbb{F}_{R_s,R_t}$ is the SO(9) crossing kernel. It is a 6j symbol, see \cite{LinZheng1} for more details. For the case above, all external operators are in the {\bf 16} irrep, e.g., the spinor irrep. Concretely, since there are 5 irreps that appear in \eqref{cyclicity4f}, $\mathbb{F}_{R_s,R_t}$ is just a $5 \times 5$ matrix with rational numbers as coefficients, see \cite{LinZheng1} equation (35) for the matrix. With this crossing kernel, we can solve \nref{cyclicity4f}. This reduces the set of 5 variables $\{a_1, a_9, \cdots, a_{126} \}$ to just two variables, which for convenience we take to be $\{a_1, a_9\} $. Furthermore, positivity\footnote{Since some of the $O$'s are actually anti-Hermitian, $a_{84}$ and $a_{36}$ are negative-definite.} of these 5 variables restricts the possible values of $\{ a_1, a_9\}$ to a subset of the $\mathbb{R}^2$:
\begin{align} \label{positivity4f1}
\frac{9!}{5!} {\red a_{126}} &= \frac{1}{6144} \tr (O^{IJKL} O^{IJKL}) = \frac{1}{672} \left(-2 {\red a_1}-2 {\red a_9}+1\right) \ge 0,\\ \label{positivity4f2}
{\frac{9!}{6!}\red a_{84}} &=  \frac{1}{1536} \tr (O^{IJK} O^{IJK}) = \frac{10 {\red a_1}-18 {\red a_9}-5}{1008} \le 0,\\ \label{positivity4f3}
{\frac{9!}{7!}\red a_{36}} &= \frac{1}{512}\tr (O^{IJ} O^{IJ}) = \frac{1}{48} \left(2 {\red a_1}+6 {\red a_9}-1\right) \le 0,\\ \label{positivity4f4}
& 9 {\red a_9} = \frac{1}{256}  \ev{\tr O^I O^I} \ge 0, \quad \left(
\begin{array}{cc}
 1 & 8 \\
 8 & 256 {\red a_1} \\
\end{array}
\right) \succeq 0  
\end{align}
In the matrix inequality \nref{positivity4f4}, we used the fact that the identity is also an SO(9) singlet, and that $\ev{\tr \psi_\alpha \psi_\alpha} = 8$. Using these constraints, we can maximize/minimize the value of $a_9$ to find that
\begin{align}\label{oioi}
    0 \le \frac{1}{9} \ev{\tr O^I O^I} \le \frac{64}{3}.
\end{align}
This value was used in deriving \eqref{bdXX}. Similarly, we can derive bounds on $a_1$ which leads to $1 \le \ev{\tr OO} \le 2$.

The main point is that after using group theory, we do not need to work with the explicit matrix $\mathcal{M}$ with its many possible values of indices\footnote{To be precise, we also considered constraints coming from considering the identity operator and the 2-fermion correlator. More generally, for each irrep $R$, we will need to consider all operators up to some level that transform in that irrep. For example, if $R$ corresponds to rank-2 anti-symmetric tensors, we will need to consider $\{ X^{[I} X^{J]}, X^{[I} P^{J]}, P^{[I} X^{J]}, \psi \gamma^{IJ} \psi \}$ up to level 3, where level is defined by \eqref{level}.}. Group theory boils down all the constraints from this giant matrix to just some simple constraints on 2 unknowns which control the values of all singlet correlators. 

\subsubsection{Numerics and other related models}
To go beyond simple analytic bounds, one can systematically automate the computation \cite{LinZheng1}. To organize the bootstrap, it is convenient to assign a level 
\begin{align}\label{level}
    \ell(X) = 1, \quad \ell(P) = 2, \quad  \ell(\psi) = 3/2.
\end{align}
Then the level of a single trace operator is the sum of the level of each letter, e.g., $\Tr X^I X^J X^I X^J$ is a level 4 operator. This assignment has the nice property that anti-commuting an operator with the supercharge \eqref{eq:susy} (at most) increases its level by 1/2. It also plays nicely with cyclicity, which either preserves the level or generates an operator with level $\ell-3$.

\begin{figure}
    \centering
    \includegraphics[width=0.9\linewidth]{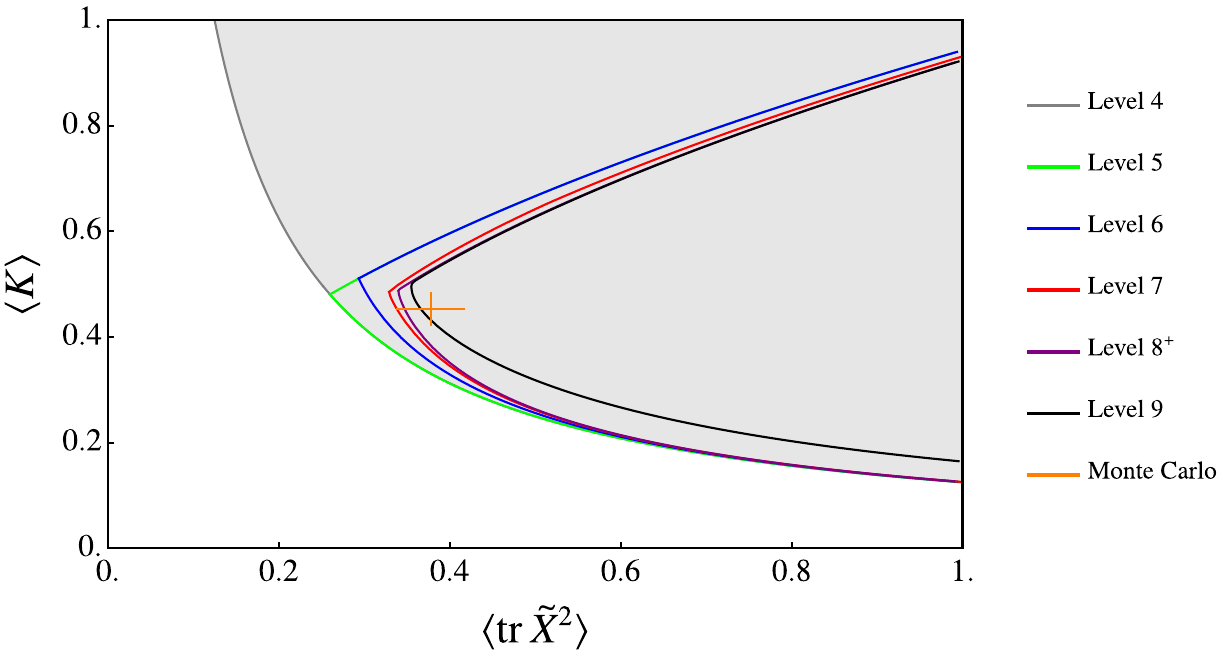}
    \caption{Bootstrap bounds for the BFSS ground state. This figure is reproduced from \cite{LinZheng1}. The $y$-axis is the kinetic energy in 't Hooft units. The cross is an extrapolation of the Monte Carlo result of \cite{Pateloudis:2022ijr}. %
    }
    \label{fig:BFSS}
\end{figure}

\begin{figure}
    \centering
    \includegraphics[width=0.9\linewidth]{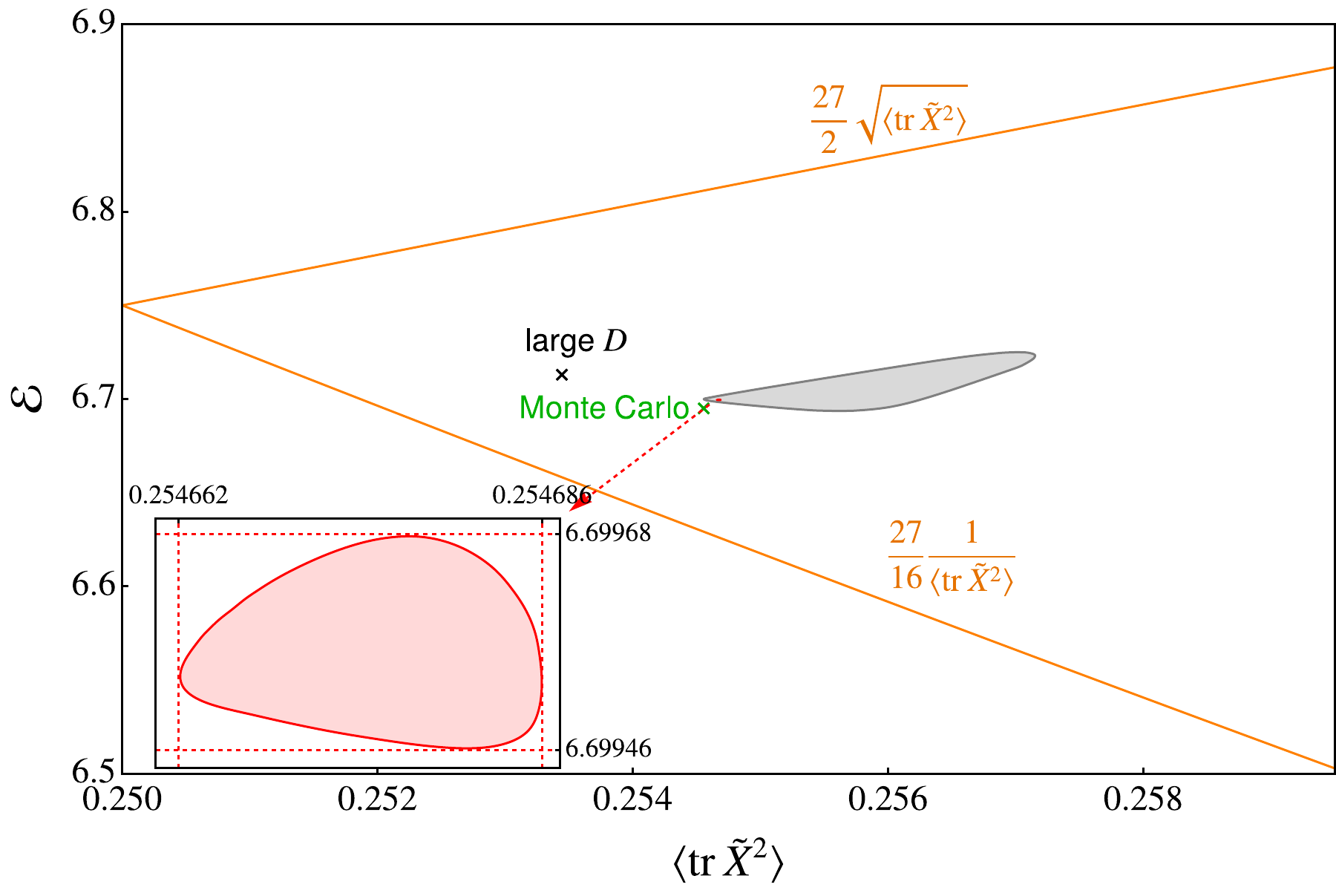}
    \caption{Bootstrap bounds for the massless $D=9$ bosonic Yang Mills matrix model, sometimes referred to as ``bosonic BFSS''. This figure is reproduced from \cite{Lin:2025srf}.
    The {\color{orange} orange analytic bound}  resembles the BFSS peninsulas displayed in Figure \ref{fig:BFSS}. At level 10 there is an island; the {\color{red} level 11 island} is barely visible; it is displayed more clearly in the inset panel.  The cross $\times$ indicates the estimate from the $1/D$ expansion \cite{Mandal:2009vz} and the {\color{dgreen} green $\times$} is from the $N=32$ Monte Carlo simulation in \cite{Kawahara:2007fn}, see also \cite{Filev:2015hia}. }
    \label{fig:bosonicBFSS}
\end{figure}

It was reported that up to level 9 the allowed region has the shape of a peninsula, see Figure \ref{fig:BFSS}. An obvious question is whether this peninsula will collapse into an island at higher levels. \cite{Lin:2025srf} studied a purely bosonic variant\footnote{Without supersymmetry, the classically flat directions of the potential where $[X_I, X_J] = 0$ are lifted quantum mechanically and the resulting theories are gapped at finite $N$. Therefore, there are no scattering states in these models. It would be interesting to study models with less than 16 SUSYs that would still have flat directions.} of BFSS:
\begin{align}
	H &= \frac{g^2_\ym}{2} \sum_{I=1}^D \left( \Tr P^I P^I \right) - \frac{1}{g^2_\ym} \sum_{I,J = 1} ^D \Tr [X^I , X^J]^2. 
\end{align}
For both $D=2$ and $D=9$, a bootstrap island was found at level 10. The island then shrinks rapidly at higher levels. A plot for $D=9$ (sometimes called ``bosonic BFSS'') has been reproduced here; one should compare the orange peninsula in Figure \ref{fig:bosonicBFSS} with the peninsulas shown in Figure \ref{fig:BFSS}. 
Optimistically one will find precise estimates by going to level 10 and beyond in the BFSS case.

\section{Conclusion}
The BFSS conjecture and, more generally, D0 brane holography is a very rich subject; we have only managed to survey a handful of developments. We have emphasized the 10d and 11d black hole in this review, but the model should also shed light on other non-perturbative objects in string/M-theory such as M2 branes \cite{Banks:1996vh}, M5 branes \cite{Banks:1996nn, Castelino:1997rv, Berkooz:1996is, Maldacena:2002rb} and Kaluza-Klein monopoles \cite{Batra:2025ivy}. It is an important window into M-theory, and perhaps has lessons for flat space holography more generally \cite{Herderschee:2023bnc, Miller:2022fvc, Tropper:2023fjr}. 
In the future, it seems important to understand new techniques for doing strongly coupled computations in this model, especially quantities that are not protected by supersymmetry. %

\section*{Acknowledgments}
I thank Gauri Batra, Shai Chester, Yiming Chen, Victor Ivo, Silviu Pufu, Juan Maldacena, Savdeep Sethi, Stephen Shenker, Douglas Stanford, Lenny Susskind, Gustavo Joaquin Turiaci and Zechuan Zheng for useful discussions. I thank all the TASI participants for their questions and for correcting some typos in an earlier version of these notes. I am supported by a Bloch Fellowship and by NSF Grant PHY-2310429.

\appendix

\section{M-theory reminder}
M-theory in asymptotically flat space is an 11d theory of quantum gravity. It has only one (dimensionful) scale, the 11d Planck scale $\ell_p$. At energies much lower than the Planck scale $E \ll 1/\ell_p$ the theory is well-approximated by supergravity in 11d. In addition to the metric, there is a 3-form gauge field $A_{\mu \nu \rho}$. The electric sources of this gauge field are called M2 branes and the magnetic sources are M5 branes. (The 4-form field strength $F$ is dual to a 7-form $\tilde{F} = d \tilde{A}$ which couples to the 6-dimensional worldvolume.)

\begin{table}[h]
 \centering  
  \caption{Relation between gauge theory, Type IIA and M-theory parameters \label{table1}}
 \begin{tabular}{|M|M|M|}  %
 \hline 
 \text{gauge theory} &  \text{Type IIA strings}  & \text{M-theory}  \\
    \hline
   N = \text{ rank of matrices} &
      
      N =\text{ \# of D0 branes} & N = \text{ KK momentum number}  \\
 \hline
    g^2_\text{YM}   & {g_s}/{(4 \pi^2 \ell_s^3)}
       & R^3/(4 \pi^2 \ell_p^6)  \\
   \hline
  \end{tabular}
\end{table}

Let us now recall the relationship between M-theory and Type IIA.
Compactifying M-theory on a small spatial circle of size $R$ gives Type IIA with
\begin{align} \label{Rgell}
    R = g_s \ell_s = g_s^{2/3} \ell_p.%
\end{align}
This follows from the fact that D0 branes have mass $M = 1/(\ell_s g_s)$ in Type IIA. On the other hand, from the 11d viewpoint they are just gravitons with Kaluza-Klein momentum $M = 1/R$.
We can derive another relation between the M-theory/Type IIA  parameters by considering the relation between an M2 brane and a string. The Type IIA string has a tension that is (by definition) $1/(2\pi \alpha') = 1/(2\pi \ell_s^2) $. On the other hand, M-theory has membranes (M2 branes) with tension $1/(2\pi)^2 \ell_p^3$. (Here $\ell_p$ is the 11d Planck scale.) To get a string (a 1-brane), we wrap the M2 on the spatial circle of radius $R$ so the resulting string has tension:%
\begin{align} \label{tensionM2}
\frac{1}{2\pi \alpha'} =   {2 \pi g_s \ell_s}\times \frac{1}{(2\pi)^2\ell_p^3}  \RA \ell_p = \ell_s g_s^{1/3} 
\end{align}
The relations \eqref{Rgell} and \eqref{tensionM2} allow us to convert the two M-theory parameters $R,\ell_p$ to the Type IIA parameters $\ell_s$ and $g_s$.

For our purposes, the low energy Lagrangian for M-theory is
\begin{align}
    I = \frac{1}{(2\pi)^8 \ell_p^9 }\int \d^{11} x \sqrt{-g} R + \cdots
\end{align}
The relation between the M-theory 11d metric and the string frame fields in Type IIA is given by
\begin{align}
\d s_{11}^2=e^{4 \phi / 3}\left(\d z+A^\mu \d x_\mu\right)^2+e^{-2 \phi / 3} \d s_{10}^2 .
\end{align}
 Plugging in this ansatz, and using the relations \eqref{Rgell} and \eqref{tensionM2} we recover the Type IIA action \eqref{actionII}.

\bibliography{main}
\bibliographystyle{JHEP}
\end{document}